\documentclass[twocolumn,trackchanges]{aastex63}
\usepackage{amsmath,amssymb,amstext,float}

\usepackage{color}

\usepackage[all]{hypcap} 
\usepackage{aas_macros}
\usepackage{natbib}
\usepackage{fancyhdr}

\usepackage{mathtools}
\usepackage{lineno}
\shorttitle{Nanoflare Diagnostics From MHD Heating Profiles}
\shortauthors{Knizhnik et al.}
\newcommand{\beg}[1]{\begin{equation}\label{#1}}
\newcommand{\done}{\end{equation}}
\newcommand{\pd}[2]{\frac{\partial #1}{\partial #2}}

\newcommand{\vecB}{\emph{\textbf{B}}}

\newcommand{\vecv}{\emph{\textbf{v}}}

\newcommand{\unit}[1]{\hat{\textbf{#1}}}

\newcommand{\curl}[1]{\nabla\times{#1}}
\newcommand{\divv}[1]{\nabla\cdot{#1}}
\newcommand{\dem}{\mathrm{EM}(T)}
\numberwithin{equation}{section}

\begin{document}

\title{Nanoflare Diagnostics From Magnetohydrodynamic Heating Profiles}

\author[0000-0002-2544-2927]{K. J. Knizhnik}
\affiliation{Naval Research Laboratory, 4555 Overlook Avenue SW, Washington, DC 20375, USA}
\author[0000-0001-9642-6089]{W. T. Barnes}
\affiliation{National Research Council Research Associate Residing at Naval Research Laboratory, 4555 Overlook Avenue SW, Washington, DC 20375, USA}
\author[0000-0003-4739-1152]{J. W. Reep}
\affiliation{Naval Research Laboratory, 4555 Overlook Avenue SW, Washington, DC 20375, USA}
\author[0000-0002-5871-6605]{V. M. Uritsky}
\affiliation{Heliophysics Science Division, NASA Goddard Space Flight Center, 8800 Greenbelt Rd, Greenbelt, MD 20771, USA}
\affiliation{Catholic University of America, 620 Michigan Ave NE, Washington, DC 20064, USA}
\correspondingauthor{K. J. Knizhnik}
\email{kalman.knizhnik@nrl.navy.mil}

\begin{abstract}
The nanoflare paradigm of coronal heating has proven extremely promising for explaining the presence of hot, \textcolor{black}{multi-million }degree loops in the solar corona. In this paradigm, localized heating events \textcolor{black}{supply} enough energy to heat the solar atmosphere to its observed temperatures. Rigorously modeling this process, however, has proven difficult, since it requires an accurate treatment of both the magnetic field dynamics and reconnection as well as the plasma's response to magnetic perturbations. In this paper, we combine fully 3D magnetohydrodynamic (MHD) simulations of \textcolor{black}{coronal active region plasma} driven by photospheric motions with spatially-averaged, time-dependent hydrodynamic (HD) modeling of coronal loops to obtain physically motivated observables that can be quantitatively compared with observational measurements of active region cores. We take the behavior of reconnected field lines from the MHD simulation and use them to populate the HD model to obtain the thermodynamic evolution of the plasma and subsequently the emission measure distribution. We find the that the \textcolor{black}{photospheric driving of the MHD model produces \textcolor{black}{only very low-frequency nanoflare heating which cannot account for the full range of active region core observations as measured by the low-temperature emission measure slope. Additionally,} we calculate the spatial and temporal distributions of field lines exhibiting collective behavior, and argue that loops occur due to random energization occurring on clusters of adjacent field lines.} 
\end{abstract}

\keywords{Sun: corona -- Sun: nanoflares -- Sun: magnetic fields}

\section{Introduction}\label{sec:intro}
The solar corona is well known to be comprised of extremely hot, thin magnetic loops that are driven at the photospheric level by convective motions. The fact that \textcolor{black}{temperature tends to increase, rather than decrease, with height along these loops was an important discovery following the measurement of coronal temperatures by \citet{Grotrian39}. Explaining the energy source for this temperature increase has been a major problem in solar physics}. The energy losses measured in active regions would cool the corona extremely quickly if there was no additional energy being added to the corona \citep{Withbroe77}. One suggested energy source for coronal heating is photospheric motions, which add stress and energy into the coronal magnetic field\textcolor{black}{. The magnetic field} then reconnects and converts its stored magnetic energy into heat as a series of localized, rapid heating events, termed `nanoflares' \citep{Parker83,Parker88,Klimchuk06,Klimchuk15,Knizhnik18a,Knizhnik19,Knizhnik20}. \par 
The details of coronal heating are much more complex than this simple picture would suggest. In reality, the frequency of nanoflares \textcolor{black}{on individual field lines} is of tremendous importance in understanding the \textcolor{black}{underlying heating mechanism} and therefore in reproducing the observed temperatures. On the one hand, if nanoflares occur constantly, almost continuously, the coronal plasma will not have time to cool, producing a nearly isothermal temperature distribution in the corona, in direct contrast to observations, which show a temperature distribution sharply peaked around $4\;\mathrm{MK}$, but having a narrow width \citep{Winebarger11,Warren12}. On the other hand, if nanoflares occur infrequently, then the distribution of temperatures in the corona will be far too broad match the observed distribution \citep{Bradshaw12,LF15}. The nanoflare frequency must be of order the coronal loop cooling time, about $10^3\;\mathrm{s}$ \citep{Cargill14,Cargill15,Klimchuk15}, \textcolor{black}{such that the plasma is allowed cool sufficiently, but still maintain a relatively narrow distribution of temperatures.} \par
The difficulty is that these results, obtained via hydrodynamic (HD) models, require the temporal and energy distribution of events which heat the plasma to be an \emph{ad hoc} input to the models because \textcolor{black}{these HD models do not self consistently model the interaction between the magnetic field and the plasma}. Thus, these HD models vary the nanoflare temporal and energy distributions until they reproduce observations, but in fact these \textcolor{black}{heating parameters} are \textcolor{black}{often} not physically motivated. One way these inputs can be determined is from fully magnetohydrodynamic (MHD) models of the solar corona driven by photospheric motions. In these models, individual field lines can reconnect and release their energy, providing a physical mechanism consistent with the imposed driving. If driving timescales are realistic, and the Poynting flux injected by the driving motions is comparable to measured values, then the time and energy scales of the resulting reconnection should be representative of the true solar values. Such MHD simulations are routinely performed to study various aspects of coronal heating \citep{Rappazzo15,Rappazzo17,Knizhnik19}.  Unfortunately, the large separation of time and spatial scales between the localized reconnection and global heating processes prohibit solving the full MHD equations, including all thermodynamic terms, at once without certain unphysical assumptions. Thus, MHD simulations are extremely useful for understanding the behavior of stressed magnetic fields and volumetric heating, but are not as useful in understanding the flow of energy between the different layers of the solar atmosphere. \textcolor{black}{In particular, it is exceedingly difficult, in practice, to identify and localize a single reconnection event and determine the energy partition (\emph{i.e.,} magnetic, kinetic, thermal) in the region before, during, and after the event. By tracing field lines, it can relatively easily be determined that reconnection has occurred, but accurately identifying where it occurred depends on the cadence of the field line tracing, which is often far less frequent than the time scale of reconnection. Furthermore, the mixing of different temperature plasmas on two reconnecting strands changes the thermal energy on each strand without contributing to the total heating \citep{Klimchuk15}. Thus, even if a reconnection event is identified and localized, determining the amount of heating that occurs from that event is extremely challenging. The only quantity that can be determined in a straightforward manner is the global heating from all of the events in the simulation.}\par 
As a result, the situation is one in which MHD simulations are able to accurately model the reconnection driving the coronal heating, but are unable to quantify the amount of \textcolor{black}{localized} heating, while HD models are able to accurately \textcolor{black}{model the thermodynamic response of the plasma}, but require an accurate description of the underlying \textcolor{black}{heating mechanism} to do so. In the absence of computational power necessary to perform a full MHD treatment of the problem that includes a proper treatment of conduction and radiation, obtaining physically motivated quantities (i.e., quantities obtained directly from MHD simulations) to feed into the HD models is a promising avenue for studying the coronal heating problem. \par   
To this end, \citet{Knizhnik20} studied the question of nanoflare frequencies using MHD simulations of a driven \citet{Parker72} plane-parallel magnetic field between two plates. Their coronal magnetic field was driven by twisting motions on one plate, and held fixed and unmoving on the other plate. They identified reconnection events that they could uniquely associate with a given field line, and found that distribution of nanoflare frequencies followed a power law with a very shallow slope of about $-1$, extending down to \textcolor{black}{the simulation cadence of} about $100\;\mathrm{sec}$ between successive nanoflares. This provides a key physically motivated input for HD models. \par 
A second physically motivated input was provided in a series of papers which measured the distribution of energies of reconnection events \citep{Kanella17,Kanella18,Kanella19,Reid20}. They quantified the spatial distribution of Joule heating to sum the energy of each event and found that the energy released per event followed a power law slope of about$-1.5$. Although there is evidence that the conversion from magnetic to thermal energy occurs via viscous, rather than Ohmic, dissipation \citep{Knizhnik19}, such temporal and energy distributions - derived from MHD directly - can be used in HD models to simulate the heating that would result from MHD simulations. \par 
The alternative approach that we develop here is to `drive' HD models by taking input directly from an MHD model and using it to seed a HD model. In this approach, the HD simulation is used to model the response of the coronal plasma to behavior of individual magnetic field lines, or `strands' in the MHD simulation. As a simple way to understand this approach, suppose that an individual field in the MHD simulation is identified and tracked with time (how this can be done is described in detail in \autoref{sec:field_line_tracing}). Then (ignoring any background heating) its heating function $q(t)$ can be defined to be $0$ where there is no change in the magnetic connectivity (i.e., no reconnection), and nonzero if there is a change in its connectivity. The exact value of $q(t)$ when it is nonzero is, to a certain extent, model-dependent. For example, in the model of \citet{Cargill14}, it depends on the time since the previous nonzero value of $q(t)$. In any case, this heating function can be input directly into an HD model, to help understand the coronal plasma's response to an MHD field line's behavior. In this way, the HD model is effectively driven or physically motivated by the MHD model. \par
In this paper, we use the approach outlined above to simulate the response of coronal plasma \textcolor{black}{inside an active region} to MHD driving and reconnection. We identify and track reconnecting field lines or `strands' in the MHD simulation, and use their behavior to seed the HD simulation. We obtain heating profiles and emission measure distributions which can be compared directly with observations. By combining \textcolor{black}{the responses of} multiple `strands' \textcolor{black}{we can predict observable signatures of `loops' comprised of multiple unresolved strands}. \par 
This paper is organized as follows. \autoref{sec:model} describes the MHD and HD models and the way the codes are combined. In \autoref{sec:Results}, we describe the results of each model separately, as well as discuss the collective behavior seen in the MHD model. We discuss the implications of the results in \autoref{sec:implications}. 

\section{Numerical Model}\label{sec:model}
\subsection{The ARMS Code}
Our simulations solve the equations of ideal MHD using the Adaptively Refined Magnetohydrodynamics Solver \citep[ARMS;][]{DeVore08} in three Cartesian dimensions. The equations have the form
\beg{cont}
\pd{\rho}{t}+\divv{\rho\vecv}=0,
\done
\beg{momentum}
\pd{\rho\vecv}{t} + \divv{\left( \rho\vecv\vecv \right)} = - \nabla P + \frac{1}{4\pi} \left( \curl{\vecB} \right) \times \vecB,
\done
\beg{energy}
\pd{U}{t}+\divv{\Big\{\left(U+P+\frac{B^2}{4\pi}\right)\vecv-\frac{\vecB(\vecv\cdot\vecB)}{4\pi}\Big\}}=0.
\done
\beg{induction}
\pd{\vecB}{t} = \curl{ \left( \vecv \times \vecB \right)}.
\done
where
\beg{totenergydensity}
 U=\epsilon+K+W
\done
is the total energy density, the sum of the internal energy density
\beg{internal}
\epsilon=\frac{P}{\gamma-1},
\done
kinetic energy density
\beg{kinetic}
K=\frac{\rho v^2}{2},
\done
and magnetic energy density
\beg{magneticenergydensity}
W=\frac{B^2}{8\pi}.
\done
In these equations, $\rho$ is mass density, $T$ is temperature, $P$ is thermal pressure, $\gamma$ is the ratio of specific heats, $\vecv$ is velocity, $\vecB$ is magnetic field, and $t$ is time. We close the equations via the ideal gas equation,
\beg{ideal}
P = \rho RT,
\done
where $R$ is the gas constant. ARMS' minimal but finite numerical dissipation allows reconnection to occur at electric current sheets associated with discontinuities in the direction of the magnetic field. 
The energy \autoref{energy} is written in conservative form, so that our simulation conserves energy, and any magnetic energy lost during reconnection is converted into plasma heating, rather than being lost from the system.\par

\subsubsection{Initial and Boundary Conditions}\label{sec:initial_boundary_conditions}
We set up a model coronal field that is initially straight and uniform between two plates \citep{Parker72, Knizhnik15, Knizhnik17a, Knizhnik18a,Knizhnik19,Knizhnik20}, each representing a photospheric boundary. The simulation setup is essentially the same as the one used in \citet{Knizhnik20}, where we state the dimensionless units used to initialize the simulation, and discuss the conversion to physical units. To avoid redundancy, here we will simply state the new simulation setup in physical units, but interested readers can refer to \citet{Knizhnik20} for a discussion of converting simulation units to physical units.

\par The domain extent in $(x,y,z)$ is $[0,L_x]\times[-L_y,L_y]\times[-L_z,L_z]$, with $x$ the direction normal to the photospheric plates. We choose $L_x=2\times10^9\;\mathrm{cm}$ and $L_y=L_z=3\times10^9\;\mathrm{cm}$ and resolve the domain with a grid of size $\delta x = \delta y = \delta z \equiv \delta = 150\;\mathrm{km}$. This is \textcolor{black}{somewhat smaller than the strand widths measured by \citet{Williams20} from recent \textit{Hi-C} observations}. At both the top and bottom plates, the magnetic field is line-tied, moving in response to motions imposed at these simulated high $\beta$ photospheres. However, the motions themselves, described below, are only imposed at the bottom plate. At the top plate, all components of the velocity are set to zero, and so the magnetic field lines are constrained to not move. As has been argued in previous papers \citep{Knizhnik17b,Dahlin19,Knizhnik20}, this is fundamentally equivalent, by symmetry, to driving at both ends at half the rate of driving at one end. The currents injected into the corona dissipate via the formation and reconnection of numerous small scale current sheets \citep{Klimchuk06,Klimchuk15,Rappazzo15, Rappazzo17,Knizhnik18a,Knizhnik19}. At all six boundaries, we employ zero-gradient boundary conditions:
\beg{zerogradient}
\pd{\xi}{n} = 0, \newline
\done 
where $\xi=\rho,T,\vecv,\vecB$ and $n=x,y,z$ is the normal coordinate. The four side boundaries are all open. The initial uniform values used in our dimensionless simulation are $\rho_0=8\times10^{-10}\;\mathrm{g\;cm^{-3}}$, $T_0=150\;\mathrm{K}$, $P_0=10\;\mathrm{dyn\;cm^{-2}}$, $B_0=50\;\mathrm{G}$. Our simulation has $R=8.26\times10^7\;\mathrm{dyn\;cm\;K^{-1}\;g^{-1}}$, an initial Alfv\'en speed $v_{A0}=B_0/\sqrt{4\pi\rho_0}=5\;\mathrm{km\;s^{-1}}$ and plasma beta $\beta_0=8\pi P_0/B_0^2=0.1$. We note that our simulations are performed in ideal MHD, so that the resistivity $\eta=0$, and magnetic reconnection occurs via the small, though finite, numerical dissipation that allows connectivity changes while conserving magnetic helicity, which is crucial for any large scale coronal model \citep{Woltjer58,Taylor74,Taylor86, Berger84,Knizhnik15,Knizhnik17a,Knizhnik17b,Knizhnik18a,Knizhnik19,Knizhnik20}. We find very similar results in the case with a small, but finite explicit resistivity \citep{Knizhnik20}. \par
It is worth mentioning that the exact values of our chosen parameters are, in many cases, quite different than their real, solar values. However, from the perspective of the physics under consideration, the density, temperature, and pressure in our model are not the key factors driving the system, and their initial values are much less crucial to the model than having realistic values of the magnetic field strength, driving velocity, and plasma $\beta$. The key physical processes that determine the amount of \textcolor{black}{energization} in the corona are the interplay between the photospheric driving and coronal magnetic reconnection. To the extent that the plasma $\beta\ll 1$ and the driving is slow compared to the Alfv\'en speed, the simulation should well represent the coronal dynamics of slow photospheric driving and low-$\beta$ magnetic reconnection. As a result, we have freedom to choose plasma properties such as pressure, density, and temperature that are far from their true values in order to obtain computationally feasible and realistic values of the driving velocity, magnetic field strength, plasma $\beta$, while having reasonable spatial and temporal resolution. In fact, it is for this reason that HD models are needed to supplement the MHD models' treatment of the plasma: it is not computationally feasible to get realistic values for all of the parameters simultaneously. We will use HD simulations, described in \autoref{sec:ebtel}, to obtain quantitative results about the plasma heating in our model.  \par 
We drive the magnetic field in our MHD simulation with a photospheric velocity profile that contains $61$ vortical cells on the bottom plate \citep{Knizhnik15,Knizhnik17a,Knizhnik17b,Knizhnik18a,Knizhnik19,Knizhnik20}. Each twist cycle for each cell consists of a slow ramp-up phase, followed by a slow decline phase. Analytically, the horizontal velocity of each cell is given by 
\beg{velcell}
\vecv_\perp(t) = \unit{x}\times\nabla\chi(r,t)
\done 
with $\unit{x}$ the vertical direction, and $r$ the radial coordinate centered on each cell. We set
\beg{chi}
\chi(r,t) = \chi_0 f(t) g(r)
\done 
where 
\beg{foft}
f(t) = \frac{1}{2}\big[1-\cos(2\pi\frac{t}{\tau})\Big]
\done 
and
\beg{gofr}
g(r) = \frac{1}{6}\Big[1-\Big(\frac{r}{a_0}\Big)^6\Big]-\frac{1}{10}\Big[1-\Big(\frac{r}{a_0}\Big)^{10}\Big].
\done 
We choose $a_0=2.5\;\mathrm{Mm}$, $|\vecv_{\perp,max}|=1\;\mathrm{km \;s^{-1}}$ and $\tau=1.6\times10^4\;\mathrm{s}$, such that the peak driving velocity is $20\%$ of the Alfv\'en speed. In our simulation, $20$ rotations of each cell are used to drive the coronal field, followed by a brief relaxation phase \textcolor{black}{of about $5$ rotations}, such that the driving duration is about $4$ days \textcolor{black}{and the relaxation phase is about $1$ day}. These parameters do not strictly represent granules or supergranules, which have primarily radial, rather than rotational, flows \citep{Schmieder14}. Instead, they model small vortices in the intergranular lanes that can be estimated to be about 10\% of a supergranular radius \citep{Klimchuk15}.  \par
\begin{figure*}
\includegraphics[width=0.5\linewidth]{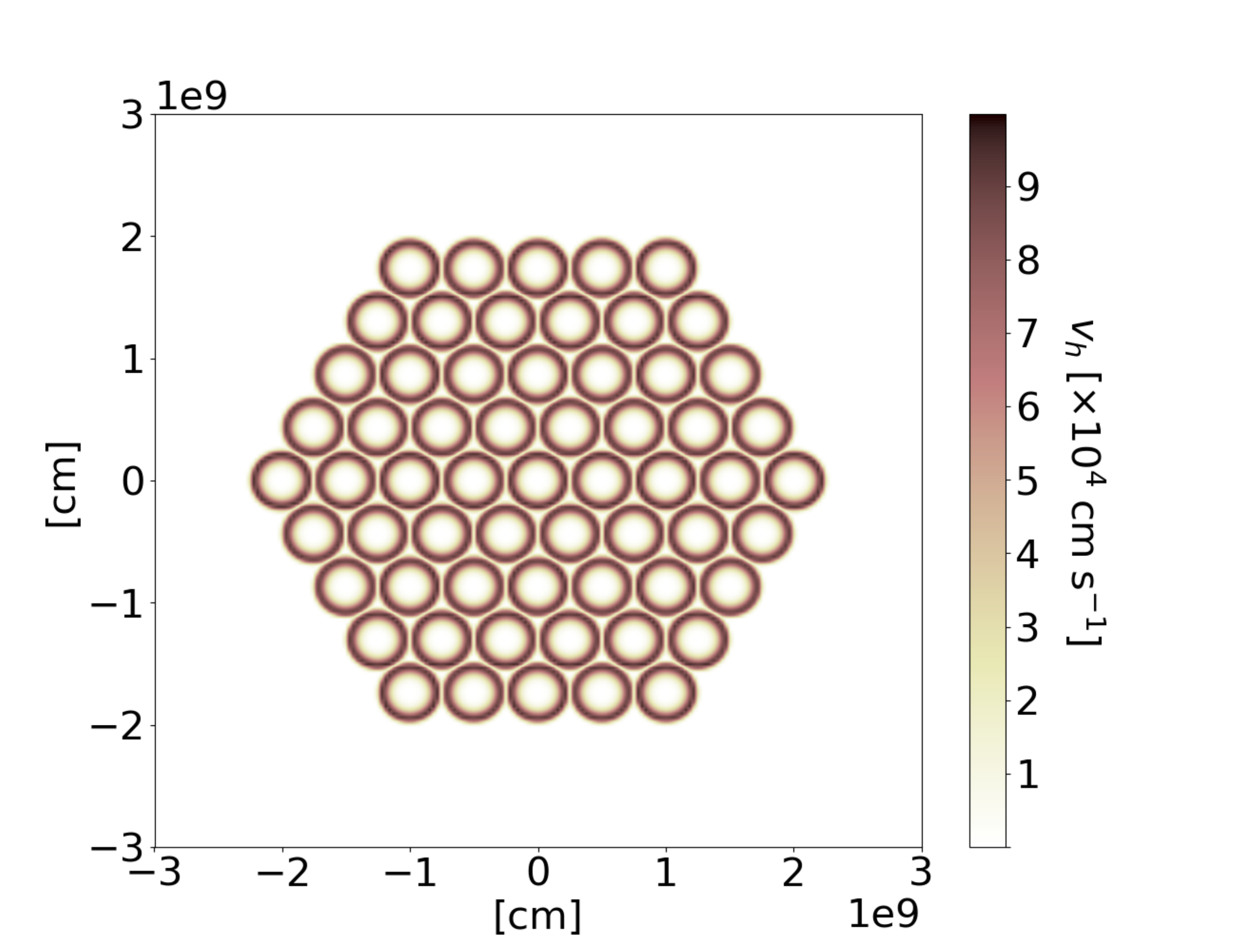}
\includegraphics[width=0.5\linewidth]{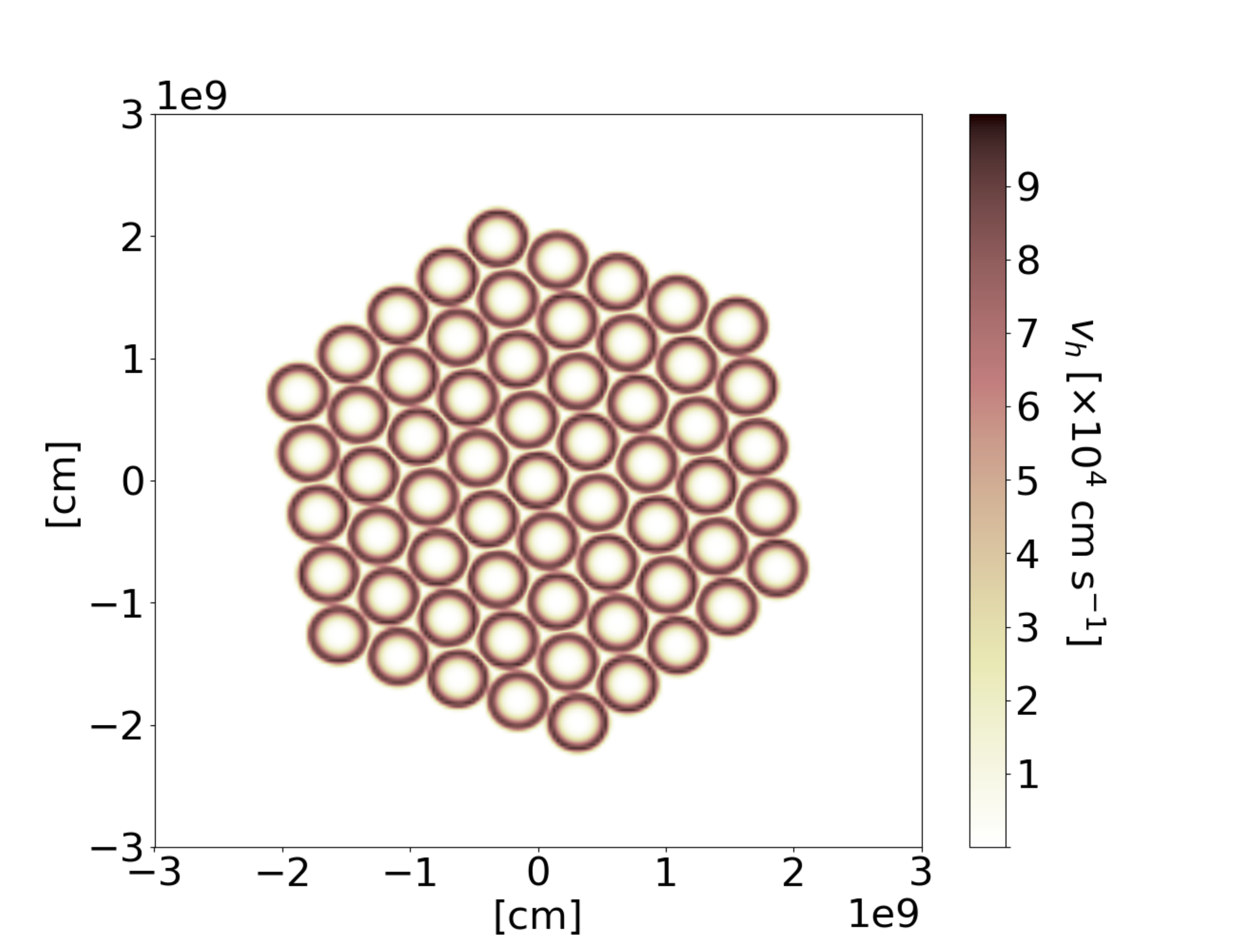}
\caption{$|V_h|$ during the first (left) and second (right) driving cycles, showing how the entire pattern is shifted after each period of rotation of the cells.}\label{fig:magV}
\end{figure*}

\begin{figure*}
\includegraphics[width=0.5\linewidth]{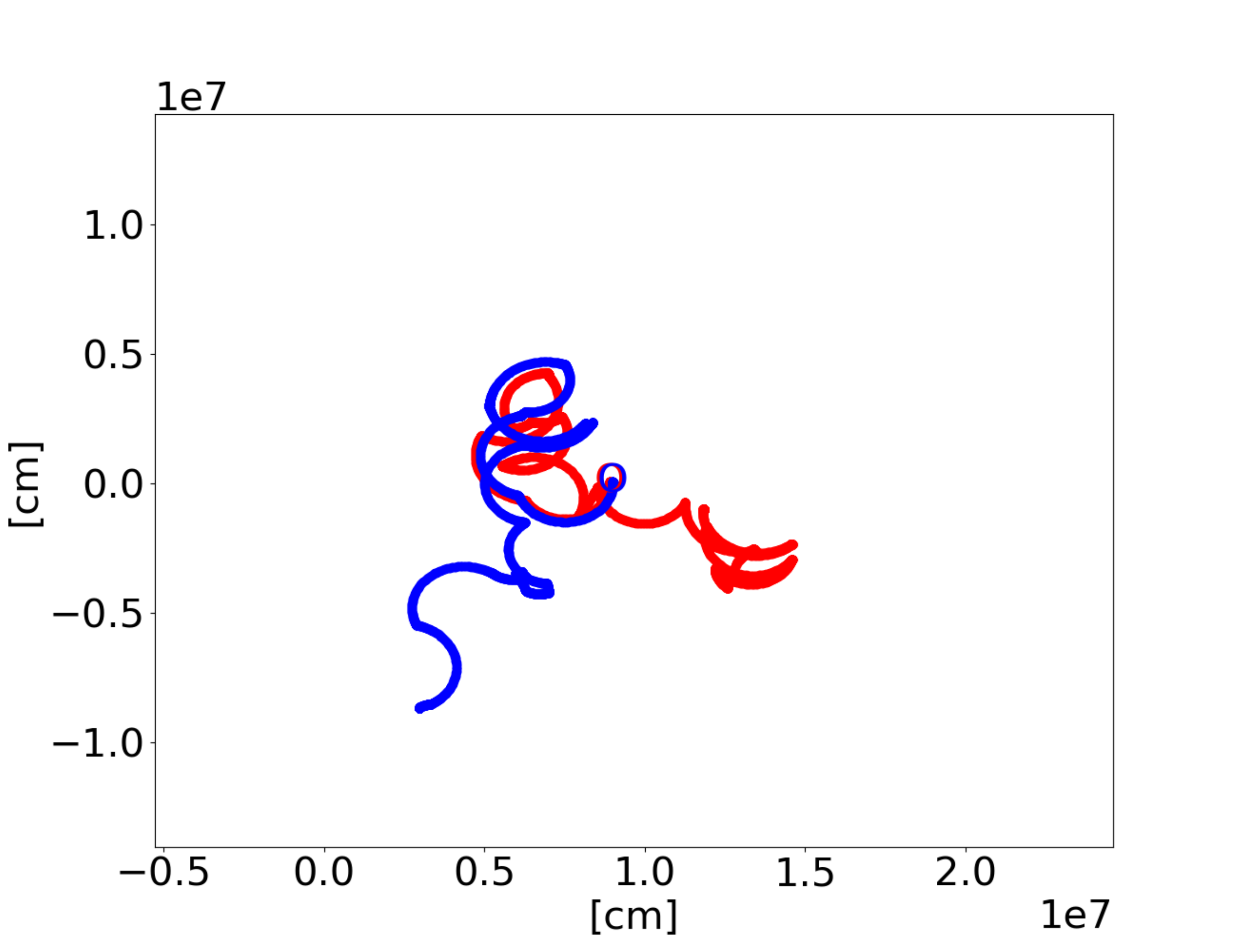}
\includegraphics[width=0.5\linewidth]{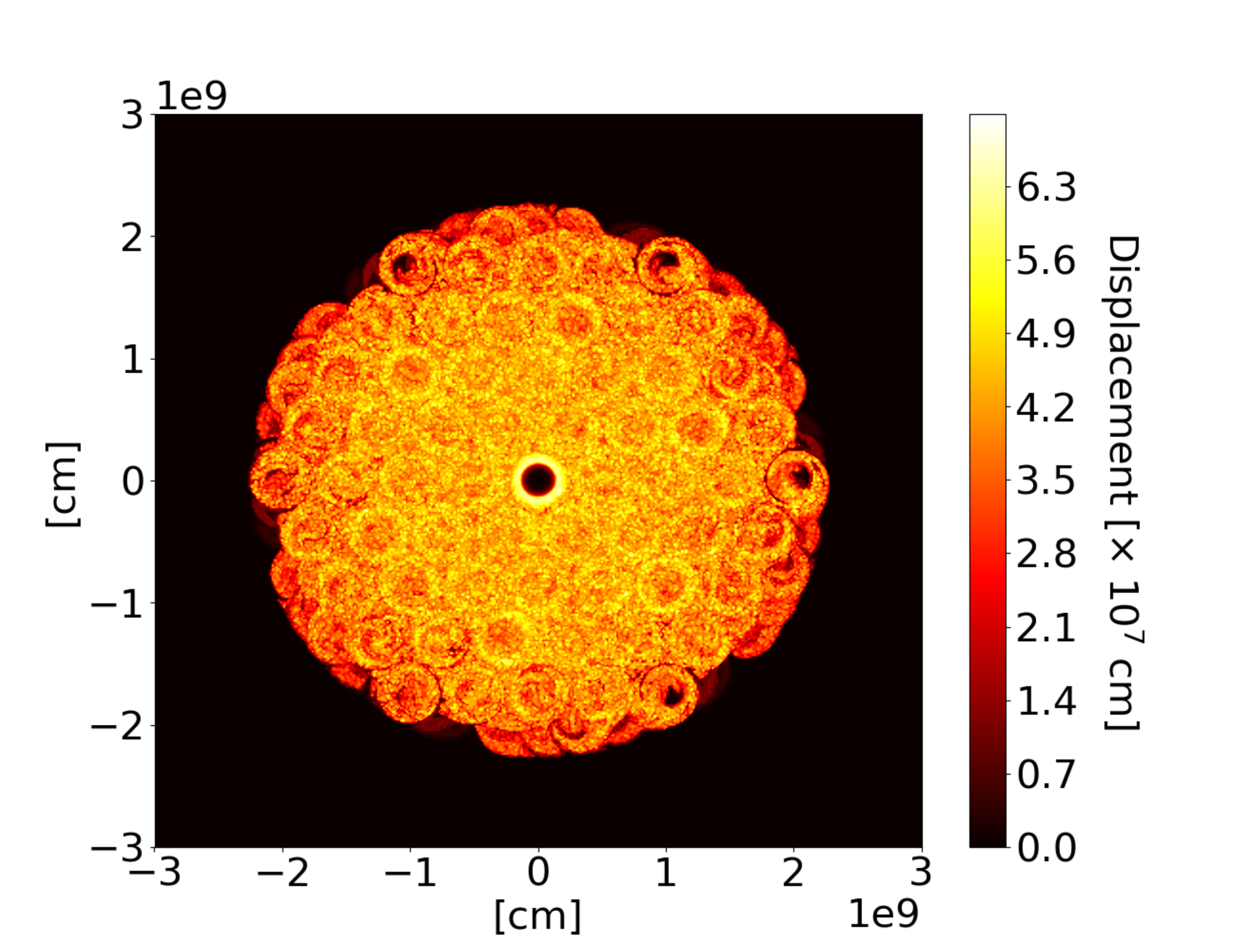}
\caption{\textbf{Left:} The path of a sample parcel of plasma on the bottom boundary that is advected by the boundary motion. \textbf{Right:} The total displacement of all parcels of plasma advected by boundary motions on a $400\times400$ grid.}\label{fig:advection}
\end{figure*}

 
The crucial difference between this work and \citet{Knizhnik20} is that we impose here a much more realistic driving pattern. Photospheric convection is expected to randomly shuffle plasma parcels around \citep{Bingert11,Rempel14,Schmieder14}, so that the motion of magnetic field footpoints should be approximately random. To increase the randomness of the driving, each rotation period of the cells is followed by an overall rotation of the entire pattern by a random angle. In \autoref{fig:magV}, we show $|V_h|$ on the bottom plane during two consecutive cycles. During each cycle, the hexagonal driving pattern is oriented at a random angle to the primary ($y-z$) axes, as was done in \citet{Knizhnik17a}. This introduces significant braiding, in addition to mere twisting, into the field line dynamics \citep{Knizhnik17a}, and allows two nearby plasma parcels to undergo significant separation. In the left panel of \autoref{fig:advection}, we plot the motion of two adjacent plasma parcels, with initial positions denoted by the red and blue `O', advected by the photospheric driving. Although the plasma parcels start out separated by one grid cell, and spend a portion of their lifetimes near each other, over the course of the simulation they move ever farther apart, allowing them to sample conditions in different parts of the domain. In the right panel of \autoref{fig:advection}, we plot the displacement of plasma parcels on the bottom boundary due to photospheric motions. With the exception of plasma parcels in the middle of the domain, which barely move due to the fact that the pattern is shifted about the center, most plasma parcels are advected distances in excess of $3-4\times10^7\;\mathrm{cm}$. In a case such as the one presented in \citet{Knizhnik20}, plasma parcels move distances of order $2\pi a_0N$, similar to the plasma near $(0,0)$ in \autoref{fig:advection}b. Thus, this simulation allows much more braiding and plasma mixing than our previous work, and results in richer field line behavior \citep{Knizhnik17a}.\par

\subsubsection{Field Line Tracing}\label{sec:field_line_tracing}

We use the ARMS field line tracer \citep{Wyper16} to trace a grid of $400\times400$ 
field lines from the fixed upper boundary at time intervals of $dt=100\;\mathrm{s}$. This is the same technique we used previously to obtain a distribution of nanoflare frequencies \citep{Knizhnik20}. It uses the fact that the top end of each field line does not move, and therefore a field line retains its identity if at two adjacent time steps $t_1$ and $t_2$, the bottom end of a field line - traced from the point $P_A$ at the top boundary - has a displacement
\beg{displacement}
d\emph{\textbf{s}}_{P^2_B-P^1_B} = \vecv_\perp(t)dt,
\done
where $P^1_B$ and $P^2_B$ are the bottom end points of the field line calculation at times $t_1$ and $t_2$, respectively. In this case, the evolution of the field is purely ideal, and the displacement is merely that due to the convective flow, \emph{i.e.}, the two field lines can be considered to be one and the same. On the other hand, if 
\beg{maxdist}
|d\emph{\textbf{s}}_{P^2_B-P^1_B}| > |\vecv_\perp(t)dt|,
\done 
then such an evolution could not have been due to the convective flow, and must have been due to reconnection. In this case, there has been a change of connectivity, and the two field lines traced from $P_A$ need to be considered as two distinct field lines. \par 
As in \citet{Knizhnik20}, we define a `nanoflare' as any event in which field lines lose their identity. In other words, if, for \textcolor{black}{any} of the $400^2$ field line seed points, the condition in \autoref{maxdist} is satisfied, we can say that a nanoflare occurred. The time interval between nanoflares, therefore, can be calculated by simply summing up the time increments $dt$ during the time interval when field lines are simply advected with the flow.  \par 
It is worth noting that our definition of reconnection in \autoref{maxdist} ignores possible reconnection events that occur on smaller scales. In other words, \autoref{maxdist} should really be an equality, and that any deviations from this equality, even displacements of $<|\vecv_\perp(t)dt|$, are due to reconnection. However, we believe that such small displacements due to reconnection events can be neglected when considering their contribution to coronal heating, since the field line is not expected to release much of its energy from infinitesimal displacements. Furthermore, due to the helicity conserving nature of our simulations \citep{Knizhnik15,Knizhnik17a,Knizhnik18a,Knizhnik19,Knizhnik20}, magnetic diffusion is negligible, and so \autoref{maxdist} will not be satisfied as a result of field line diffusion. \par 

\begin{figure*}
\includegraphics[width=0.5\linewidth]{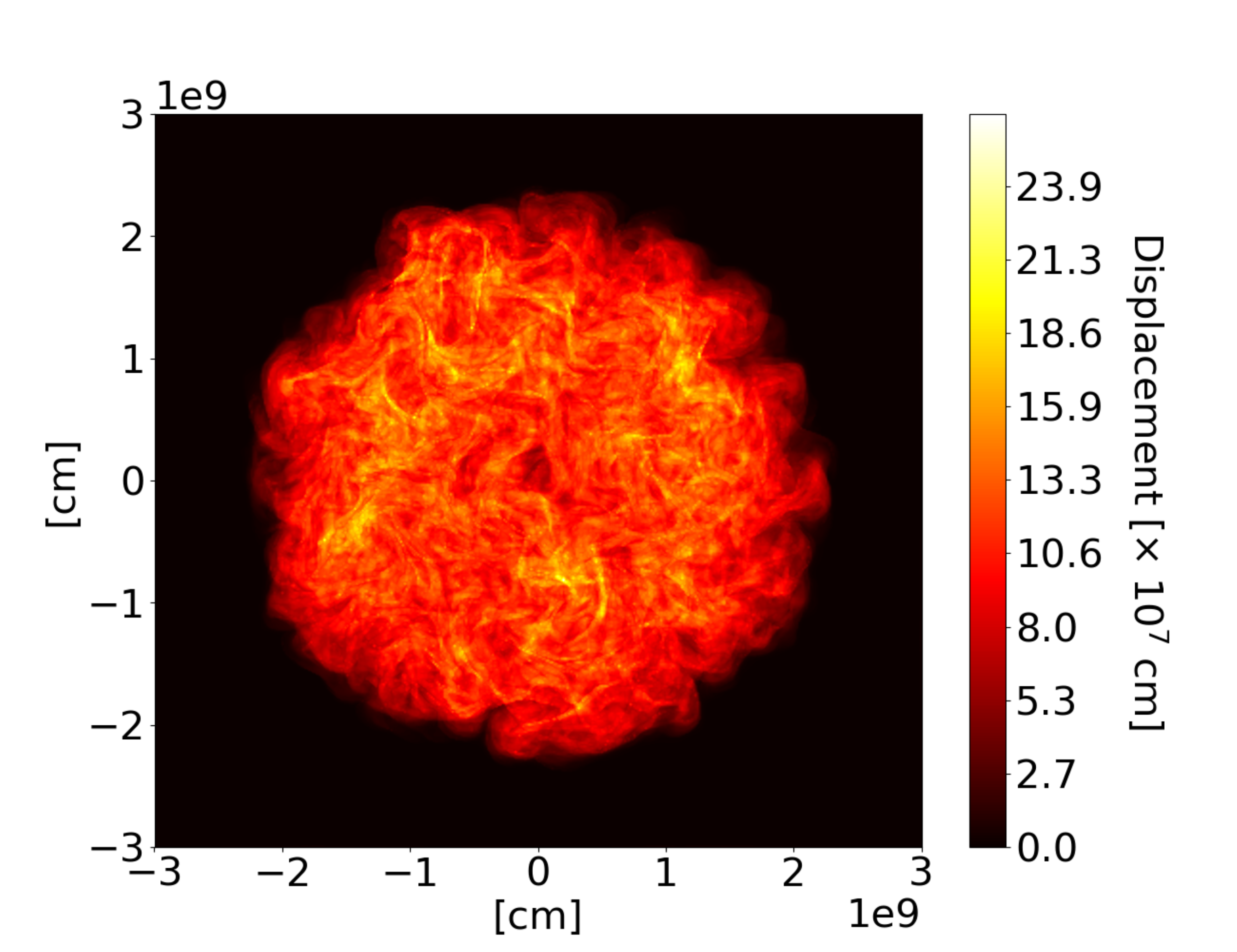}
\includegraphics[width=0.425\linewidth]{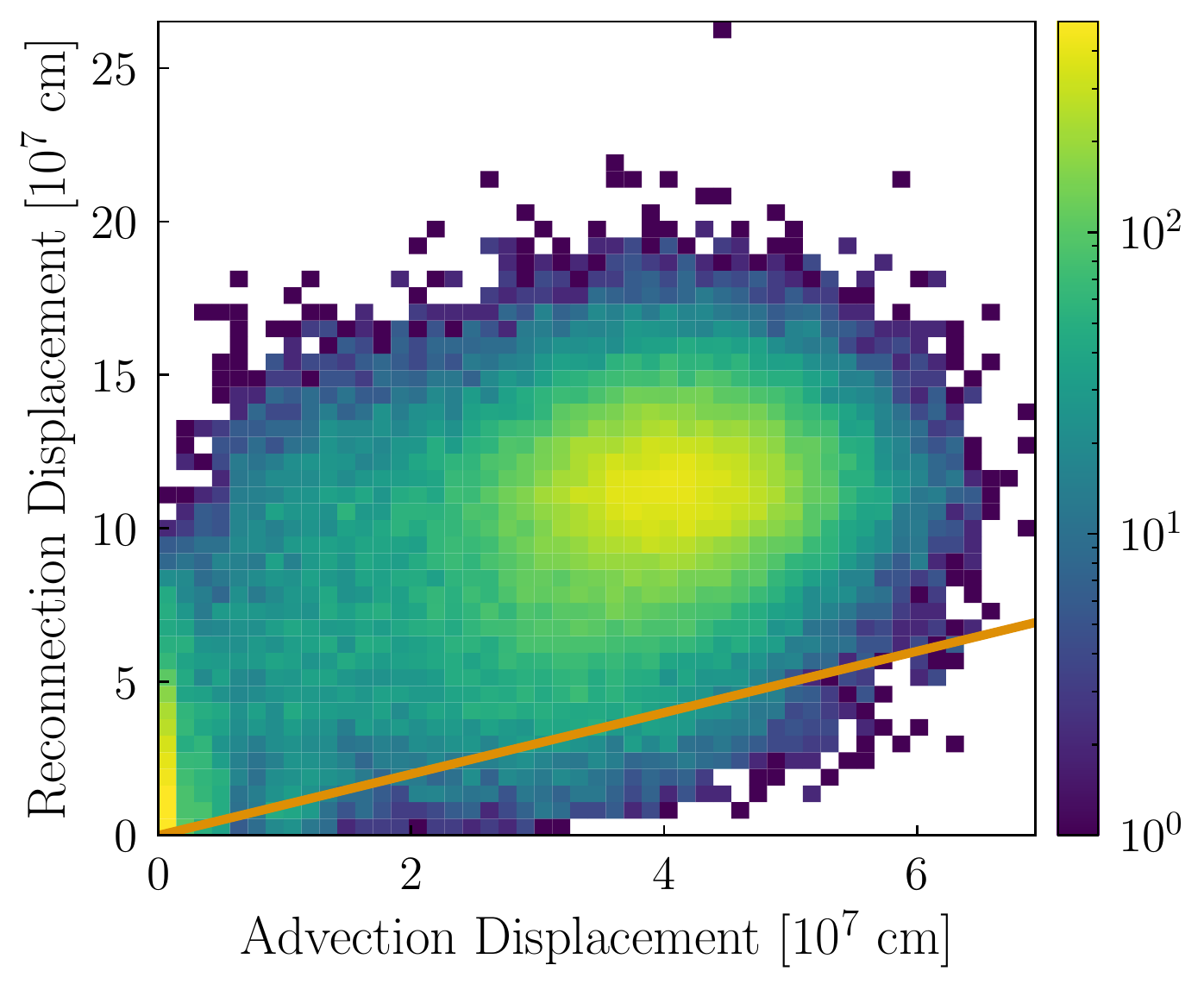}
\caption{\textbf{Left:} The total displacement the bottom footpoints of field lines traced from the top (fixed) boundary on a $400\times400$ grid. \textbf{Right:} The displacement of the bottom ends of field lines in the MHD simulation vs. displacement of plasma parcels by boundary motions. The orange line has a slope of $1$. }\label{fig:displacements}
\end{figure*}

The left panel of \autoref{fig:displacements} shows a map of displacements of the bottom ends of field lines traced from a $400\times400$ grid on the top (fixed) boundary. If reconnection were suppressed, this map would look exactly like the one in the right panel of \autoref{fig:advection}. In the presence of reconnection, however, the field lines travel nearly $3-4$ times farther than just due to photospheric motions. The right panel of \autoref{fig:displacements} shows a histogram of displacements due to reconnection and advection. The orange line, with slope unity, shows that field lines move much farther throughout the domain due to reconnecting with other field lines and changing their connectivity than plasma parcels would just due to advection. It is clear that reconnection plays a significant role in the motion of field lines in our simulation. They are not merely advected with the flow, but undergo constant changes in connectivity that displace their footpoints on size scales much larger than the convective cell diameter.\par 

\subsection{The EBTEL Code}\label{sec:ebtel}

As noted in \autoref{sec:initial_boundary_conditions}, the focus of the ARMS code is the dynamics and evolution of the magnetic field rather than the thermodynamics of the plasma. As such, we use the Enthalpy-Based Thermal Evolution of Loops \citep[EBTEL,][]{Klimchuk08,Cargill12,Cargill12b} model, specifically, the two-fluid version of EBTEL \citep{Barnes16}, to simulate the thermodynamic evolution of the coronal plasma for each traced strand, as discussed in \autoref{sec:field_line_tracing}, in response to a time-dependent heating function derived from the reconnection profiles for each strand (as described in \autoref{sec:strandVloop}). The two-fluid EBTEL code solves the time-dependent, two-fluid HD equations spatially integrated over a symmetric, semi-circular coronal strand for the coronal temperature, pressure, and density for both the electron and ion fluids. The model accounts for radiative losses from both the transition region and corona, thermal conduction (including flux limiting), and energy transfer between the electron and ion fluids via binary Coulomb collisions. The two-fluid EBTEL code is written in C++ and is very efficient, capable of computing solutions for hundreds of thousands of strands for many days of simulation time in only a few hours\footnote{The complete source code, including examples and documentation for ebtel++, the two-fluid C++ EBTEL code, are available here: \href{https://github.com/rice-solar-physics/ebtelPlusPlus}{github.com/rice-solar-physics/ebtelPlusPlus}}. A complete description and derivation of the two-fluid EBTEL equations can be found in \citet[Appendix B]{Barnes16}. \textcolor{black}{In \S \autoref{sec:ebtelsim}, we describe the specific parameters used in the EBTEL simulations and how they relate to the output parameters from ARMS.}

\subsubsection{The Emission Measure Distribution Diagnostic}
\label{sec:em_diagnostic}
\textcolor{black}{The primary diagnostic that will be considered for the analysis is the }
column emission measure distribution,
\beg{defdem}
\dem=\int\mathrm{d}h\,n_e^2\;\;\;\;\;[\mathrm{cm^{-5}}], 
\done 
where $n_e$ is the electron density and the integration is taken along the line of sight $h$. This is an often-used diagnostic for constraining the frequency of energy deposition in active region core observations. In particular, the ``cool'' portion of $\dem$ (i.e. leftward of the peak, $10^{5.5}\lesssim T\lesssim10^{6.5}$ K), can be described by a power-law relation $\dem\sim T^a$ \citep{Jordan76,Cargill94}. The \textit{emission measure slope}, $a$, so called because it corresponds to a linear slope in $\log-\log$ space, can be used to distinguish between high- and low-frequency nanoflare heating \textcolor{black}{on individual strands} \citep[see Table 3 of][and references therein]{Bradshaw12}. The emission measure slope typically falls in the range $2<a<5$, with shallower slopes indicative of low-frequency heating and steeper slopes associated with high-frequency heating \citep[e.g.][]{Tripathi11,Warren11,Winebarger11,Schmelz12,Warren12,DelZanna15}.

For each EBTEL simulation, described below, we compute the coronal $\dem$ using the resulting time-dependent electron temperature, $T$, and density, $n$, as calculated by EBTEL. At each time step $t_i$, we bin the temperature $T_i$, weighted by \textcolor{black}{$n_i^2L_x/2$ (where $L_x = 2\times10^9\;\mathrm{cm}$ is the loop length)},  into a set of bins with left and right edges at $10^5$ and $10^8$ K respectively, spaced evenly in $\log{T}$ with widths of $\Delta\log{T}=0.05$. We then \textcolor{black}{average} along the time axis to compute the time-averaged $\dem$ for the entire simulation. To compute the $\dem$ slope $a$, we apply a first-order polynomial fit to the $\log-$transform of the total time-averaged $\dem$ in each bin and the $\log-$transform of the bin center. We apply this fit only over bins in the range $10^6\,\mathrm{K}\,<T<T_M$, where $T_M$ is the temperature corresponding to the peak of the $\dem$ distribution. 

\section{Results}\label{sec:Results}

\subsection{Time Delays Between Reconnection Events}
\begin{figure*}
\includegraphics[width=0.5\linewidth]{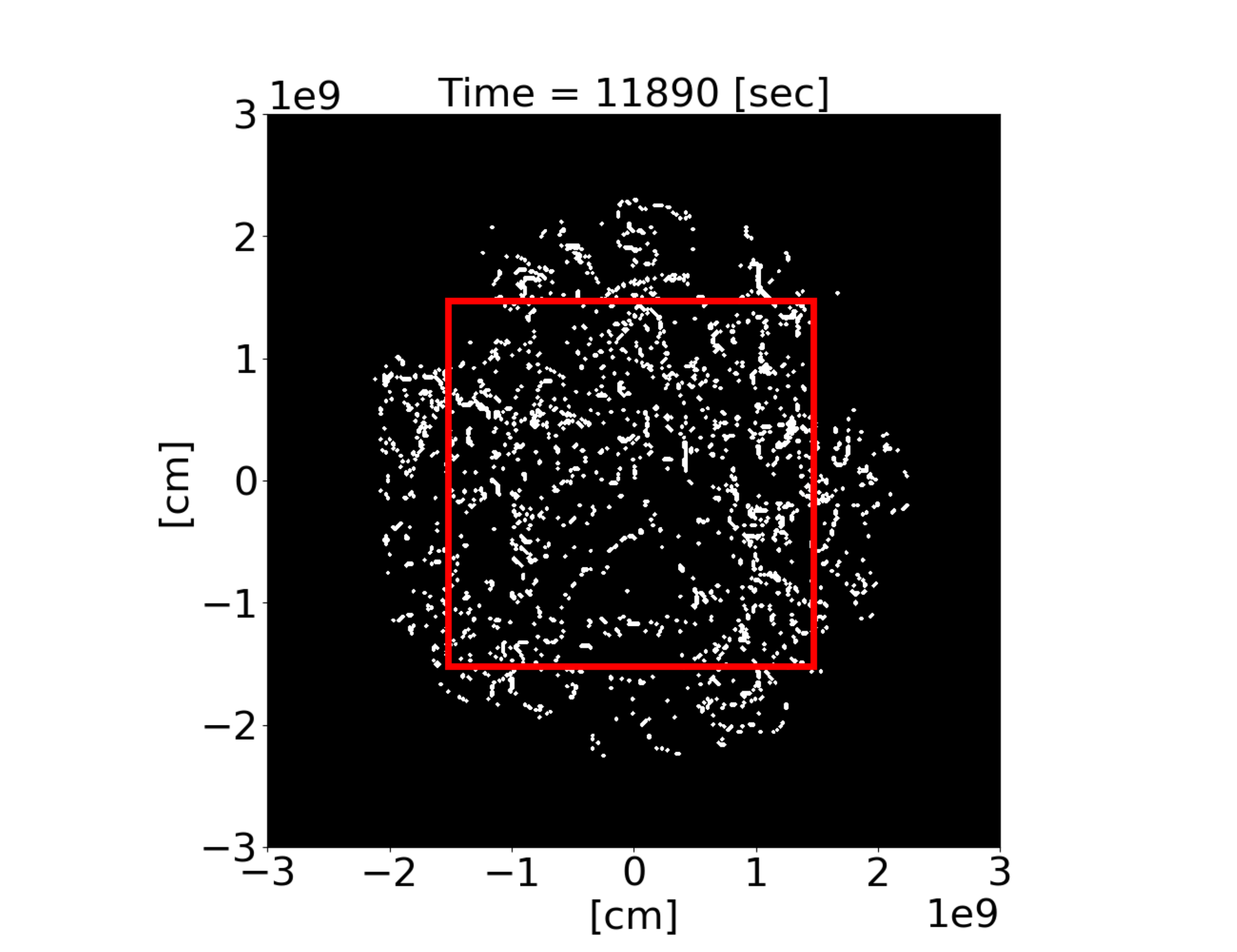}
\includegraphics[width=0.5\linewidth]{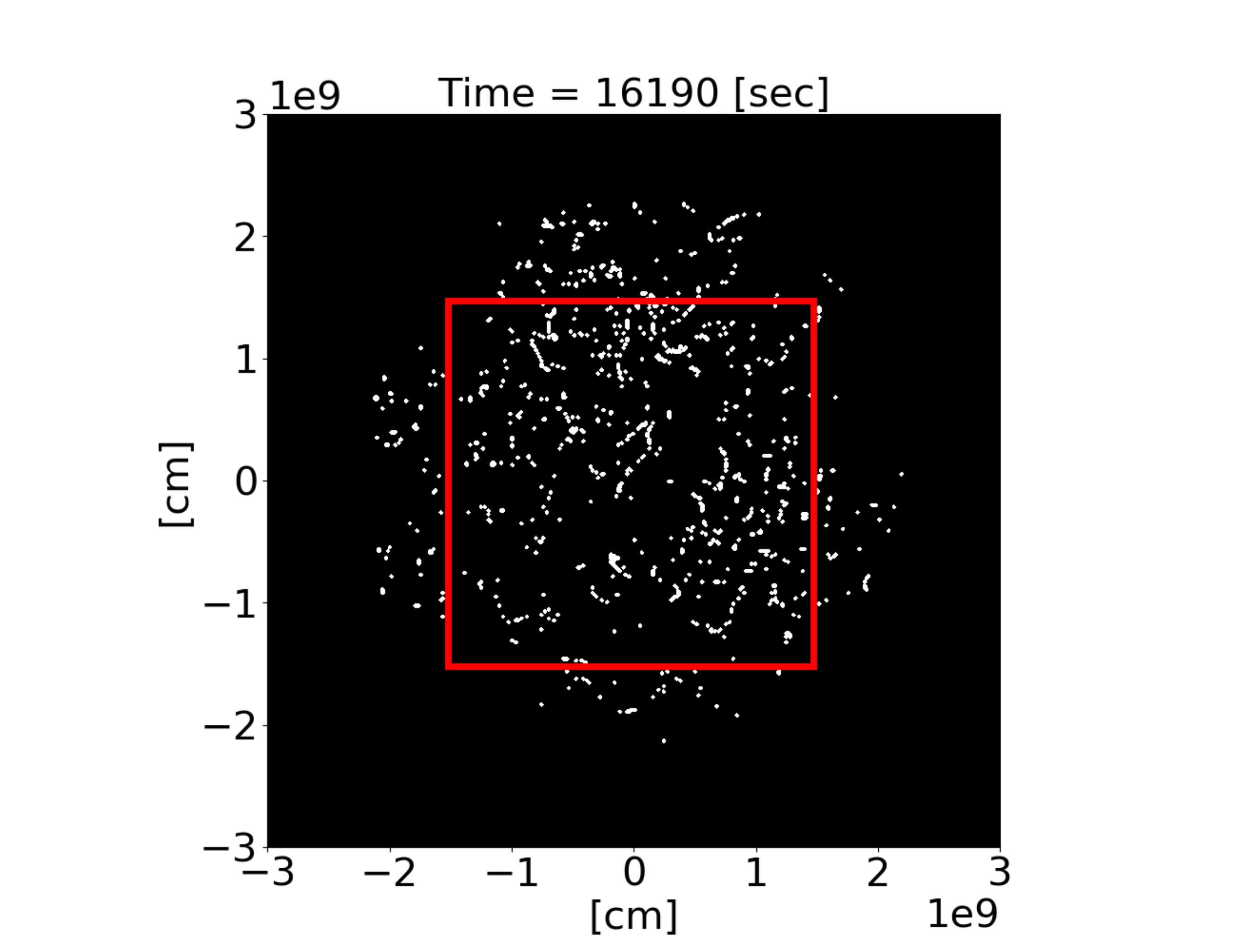}
\includegraphics[width=0.5\linewidth]{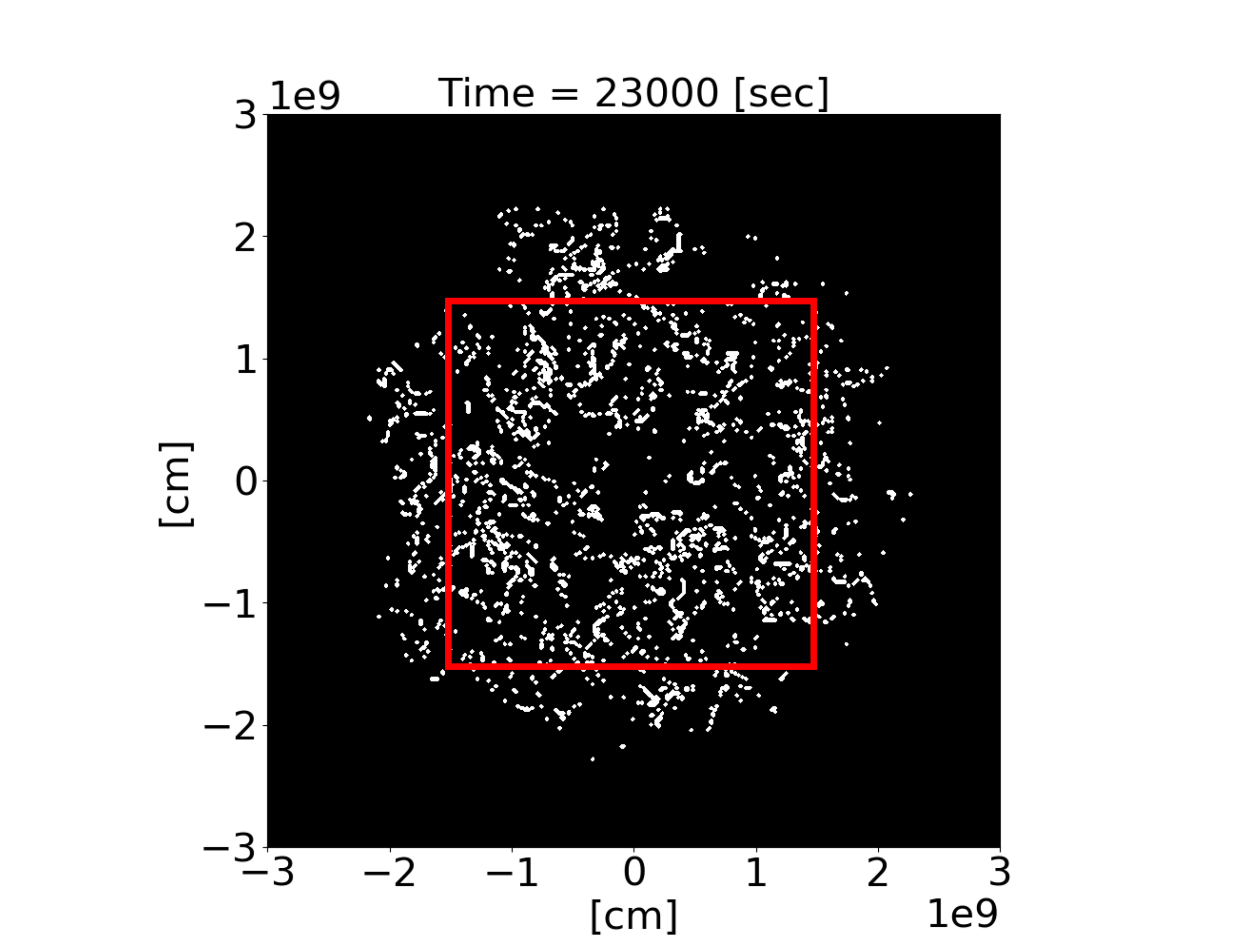}
\includegraphics[width=0.5\linewidth]{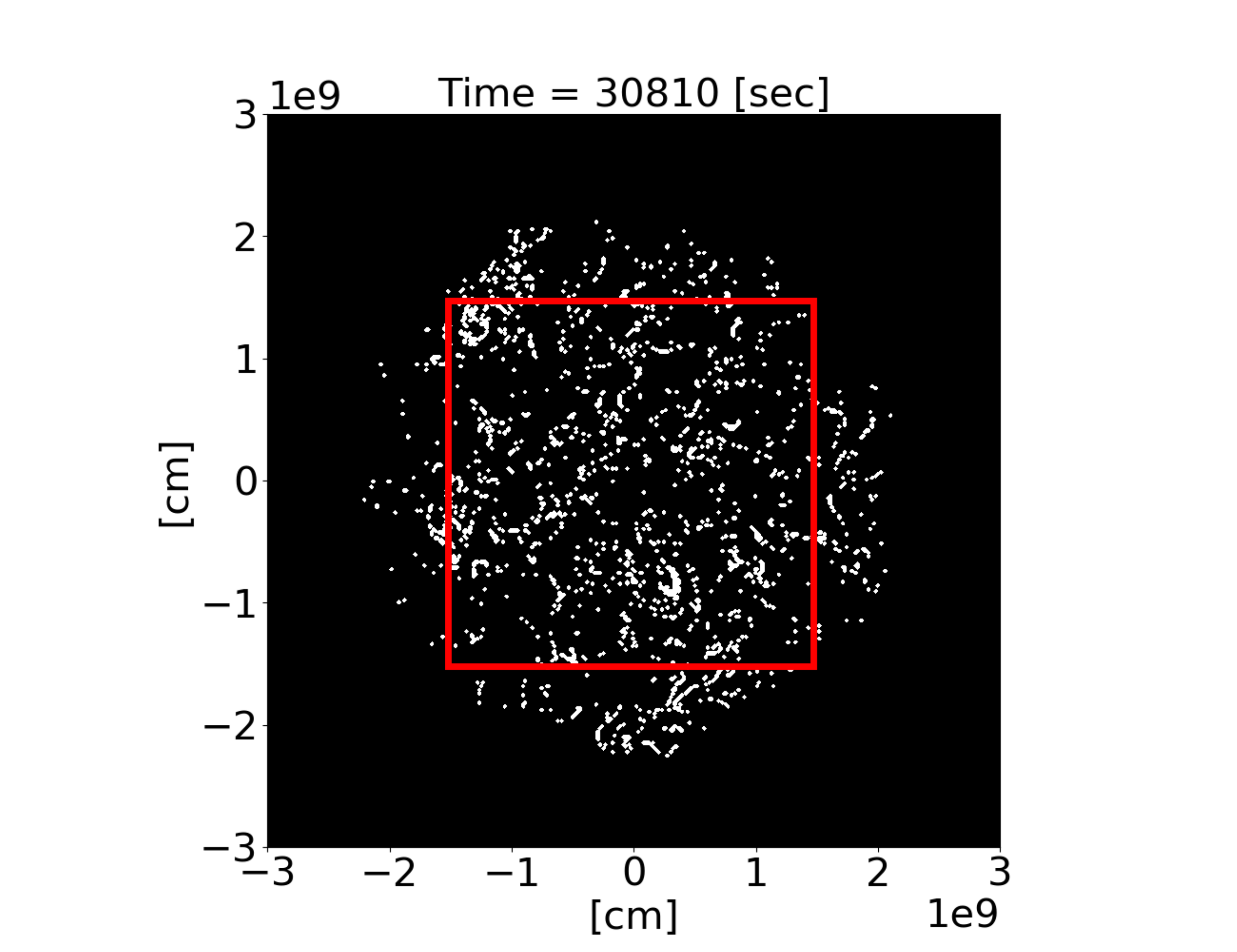}
\caption{Maps showing locations of reconnection (white) and no reconnection (black) at four different times during the simulation. The red box shows the region which was considered for the subsequent analysis.}
\label{fig:InstRx}
\end{figure*}

\begin{figure}
\includegraphics[width=\linewidth]{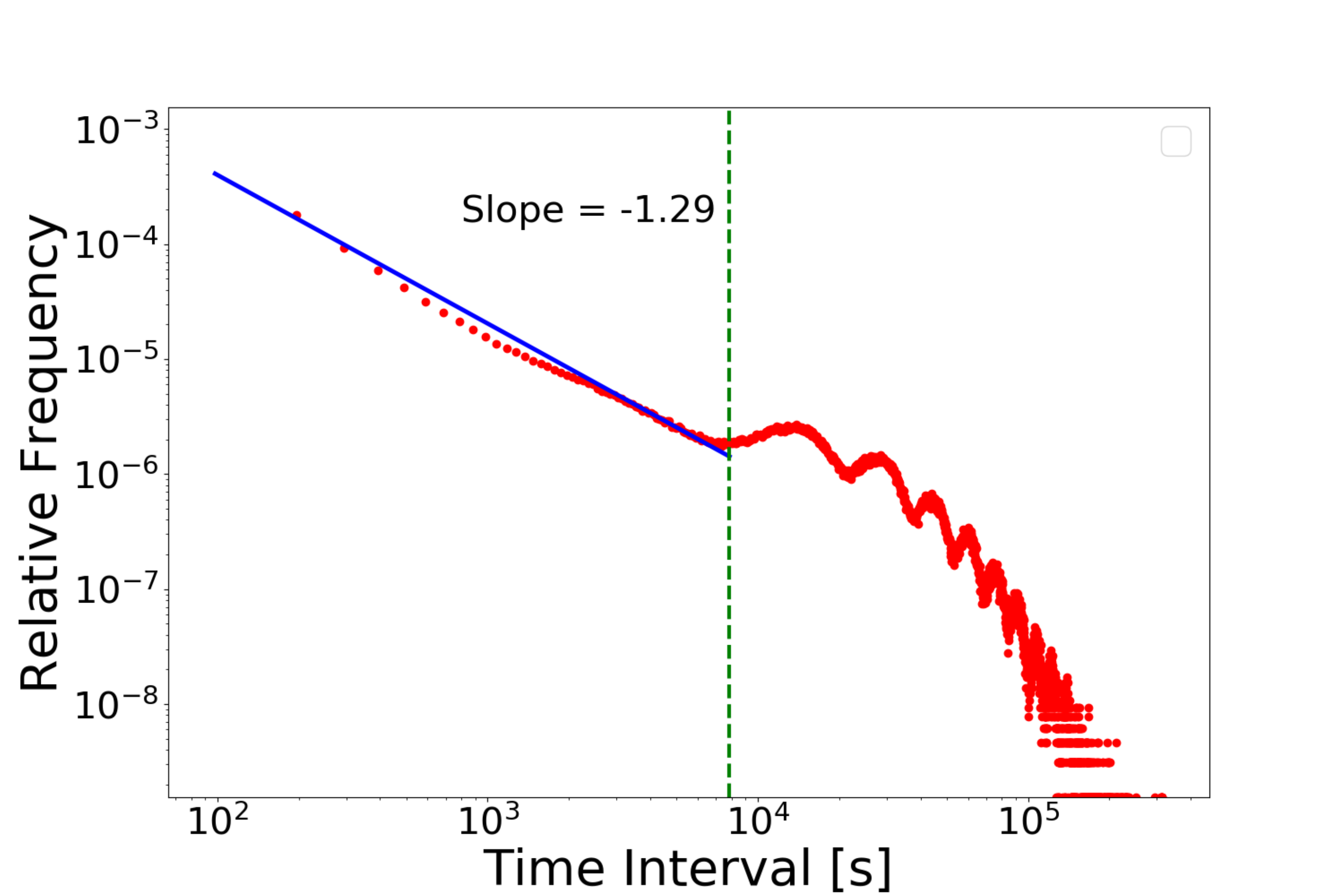}
\caption{The time delay distribution between successive reconnection events, showing a slope very similar to \citet{Knizhnik20}.}\label{fig:tdelay}
\end{figure}

We begin our presentation of the results by investigating whether the more complex driving pattern changes the distribution of nanoflare frequencies from that measured in \citet{Knizhnik20}. In \autoref{fig:InstRx}, we plot locations where field lines are instantaneously changing their connectivity at several times during the simulation. The white dots are locations on the top plate where the bottom footpoint of the field line traced from that location reconnects. During the course of the simulation, the seed points for field line tracing show up as white or black depending on whether there has been a reconnection. For each grid point, we can calculate the time between successive reconnection events, as described above. \autoref{fig:tdelay} shows a distribution of these time delays, calculated only for $201\times201$ field lines traced from within the red box, shown in \autoref{fig:InstRx}. The measured slope of $-1.29$ is slightly steeper than that found in \citet{Knizhnik20} for a fixed pattern, but is nevertheless far shallower than previous studies have assumed \citep{Cargill14,Bradshaw16,Barnes16b,Barnes19}. The green line represents the half period of rotation of the convective cells. The slightly steeper slope seen in this simulation as compared to \citet{Knizhnik20} is likely due to the fact that braiding is able to bring field lines from larger separations closer together, resulting in a higher likelihood and frequency of reconnection, especially on shorter time scales. Power-law clustering of waiting times of nanoflares is indicative of MHD turbulence, and are unlikely to result in self-organized critical systems \citep{Carbone02,Uritsky14}. \par

\subsection{Collective Behavior}
\label{sec:collective}

To understand the relevant spatial and temporal scales of our reconnection events, we utilize an event tracking algorithm \citep{Uritsky10a,Uritsky10b,Uritsky14} which identifies clusters of spatially and temporally adjacent reconnecting grid points \citep{Knizhnik18a}, i.e., it finds contiguous white regions in \autoref{fig:InstRx} in both space and time. \textcolor{black}{Thus, a `cluster' is here defined as the set of field lines traced from spatiotemporally adjacent reconnecting grid points. For example, field lines traced from three adjacent grid points which all reconnect are identified as a cluster. If, in the next time step, two of the field lines reconnect, but the third does not, it is still defined as the same cluster even if the non-reconnecting field line was in between the two reconnecting field lines, so that the two reconnecting field lines are no longer adjacent to each other. Thus, a cluster can, in theory, spatially split up while still being temporally connected.} We identify the event (linear) size,
\beg{linearsize}
L_i = \delta \max_{\forall j,k\in\Lambda_i}\left((y_j-y_k)^2+(z_j-z_k)^2\right)^{1/2},
\done 
which represent the largest spatial separation between the grid nodes included in the event, the instantaneous time-dependent event area,
\beg{area}
\mathcal{A}_i(t) = \delta^2 \sum_{k\in\Lambda_i(t)}{k},
\done 
the event lifetime based on its starting and ending times (respectively $t_s$ and $t_e$),
\beg{lifetime}
\tau_i = t_e-t_s,
\done 
the maximum event area,
\beg{area}
A_i = \max_{t\in[t_s, t_e]} \mathcal{A}_i,
\done 
and the spatiotemporal volume,
\beg{volume}
V_i = \delta t \,\, \delta^2 \sum_{k \in \Lambda_i} k.
\done 
Here, $\Lambda_i$ is the set of all spatiotemproal positions involved in the $i^{th}$ event, $\delta = \delta y = \delta z = 1.5\times10^7\;\mathrm{cm}$ is the strand width defining the spatial resolution of the studied arrays, $\delta t$ is the time step between consecutive arrays used to detect the events, $y$ and $z$ are the dimensionless grid point  coordinates, and $k$ and $j$ are the grid point labels. 

\begin{figure*}
\includegraphics[width=0.5\linewidth,trim=0.0cm 18.0cm 7.0cm 0.0cm,clip=true]{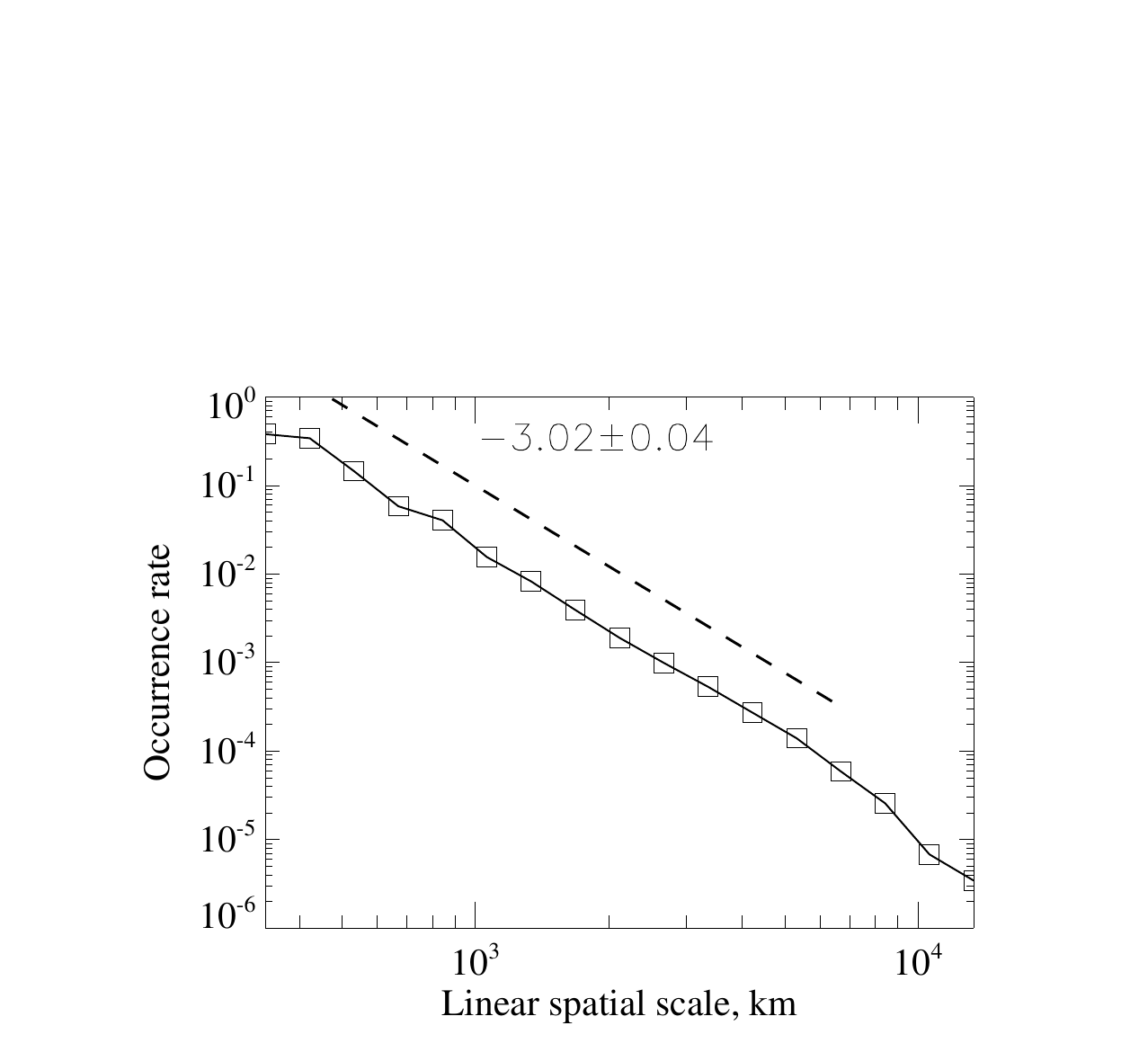}
\includegraphics[width=0.5\linewidth,trim=0.0cm 18.0cm 7.0cm 0.0cm,clip=true]{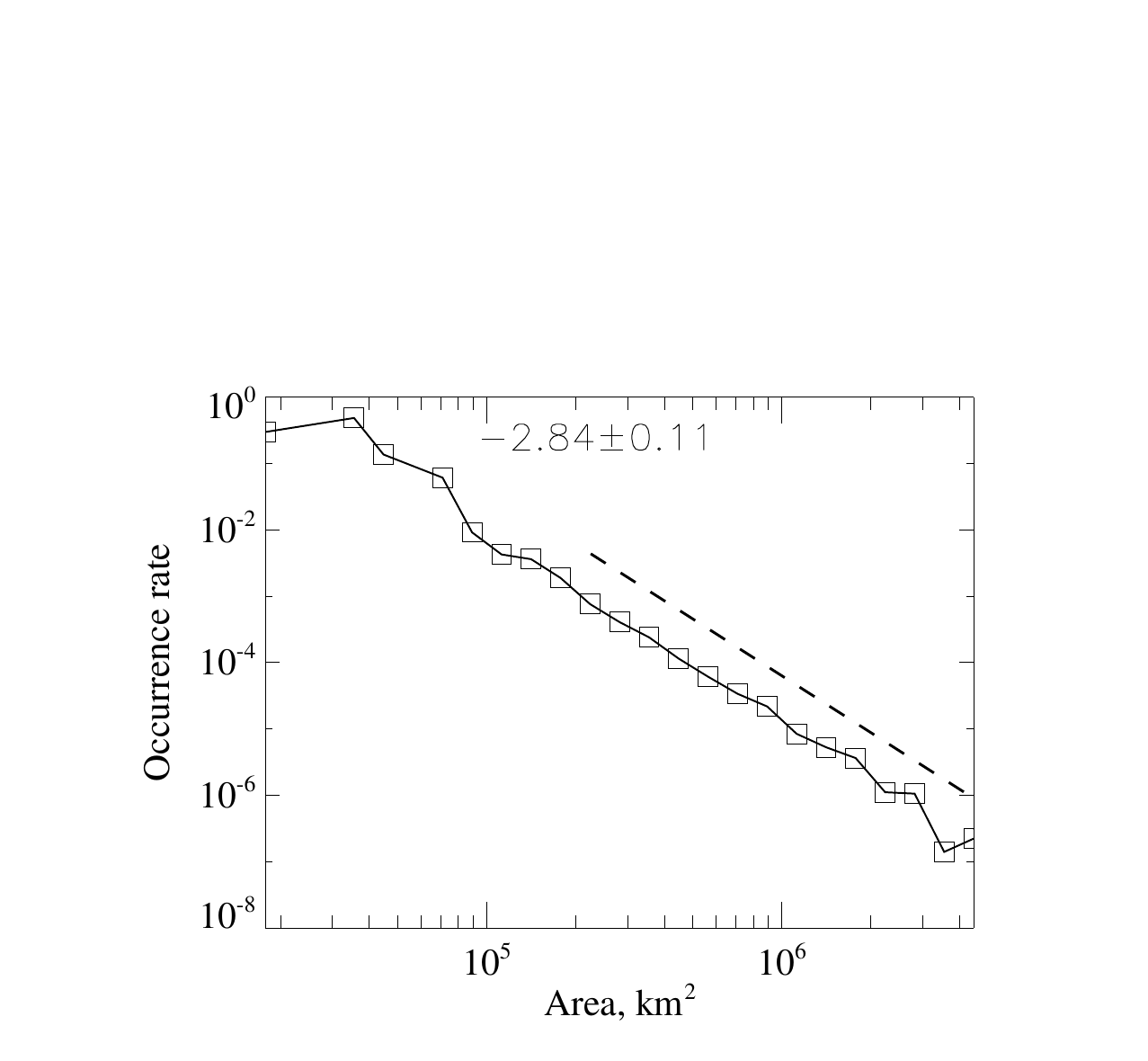}
\includegraphics[width=0.5\linewidth,trim=0.0cm 18.0cm 7.0cm 3.0cm,clip=true]{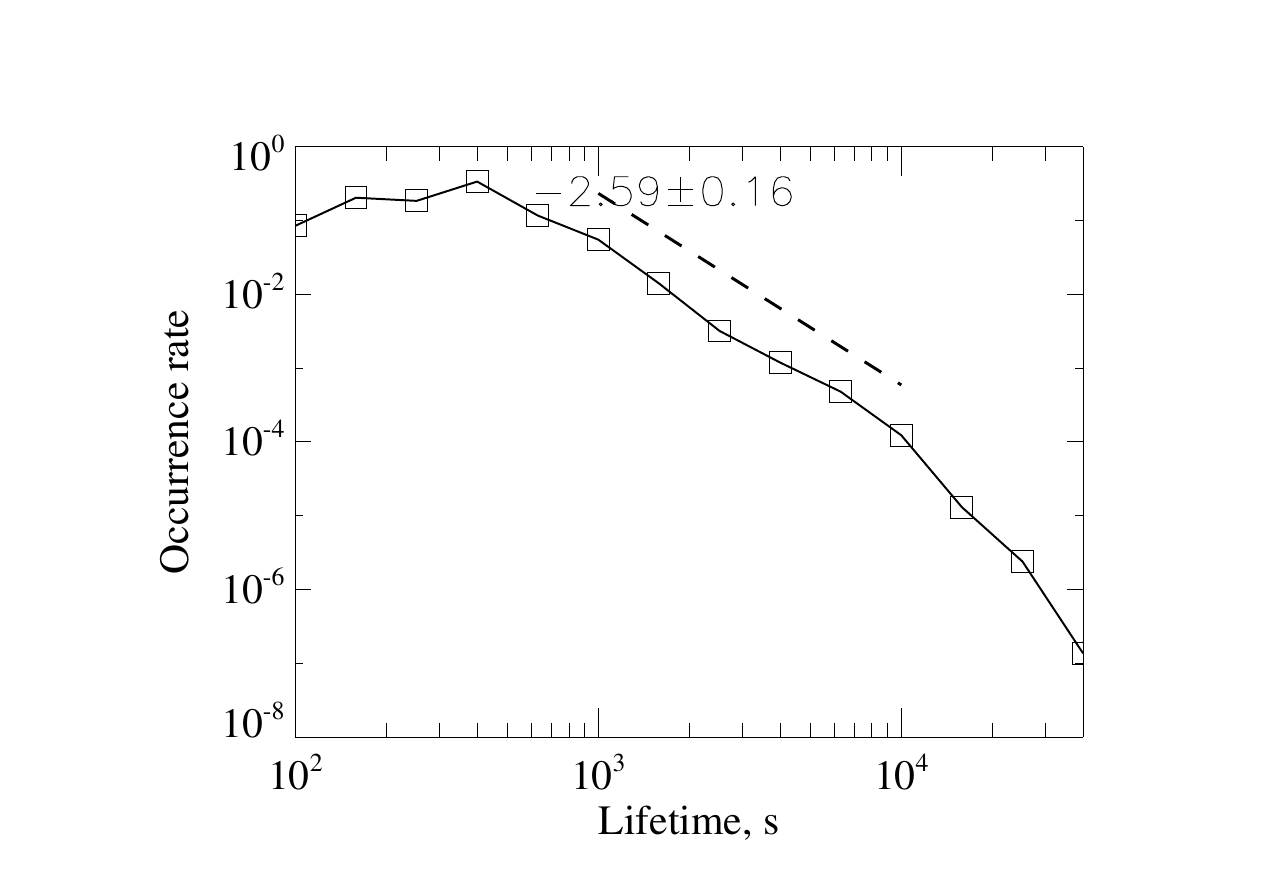}
\includegraphics[width=0.5\linewidth,trim=0.0cm 18.0cm 7.0cm 3.0cm,clip=true]{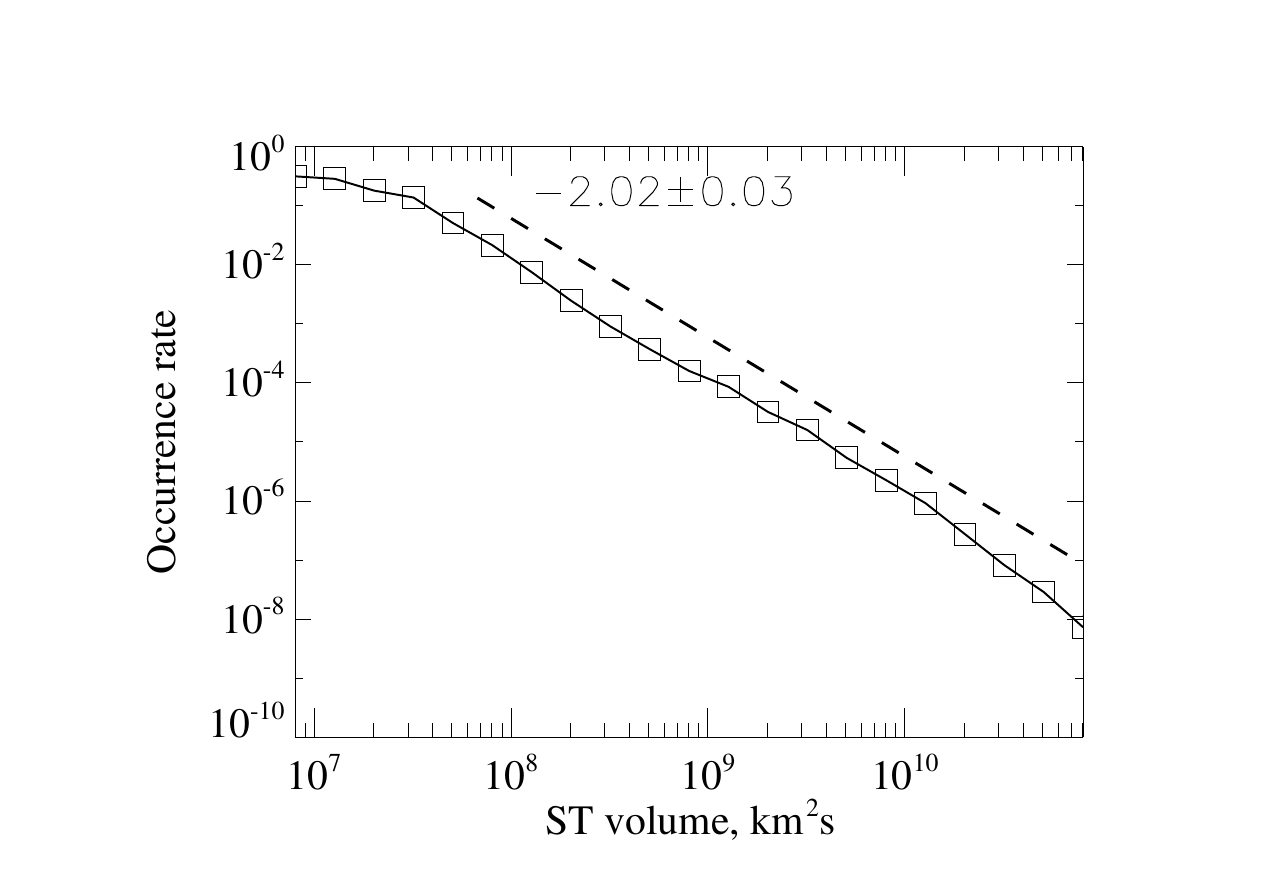}
\caption{The results of the cluster analysis, showing the probability distribution of cluster lengths, areas, lifetimes, and spatiotemporal sizes, and the associated power-law slopes accompanied by standard errors.}\label{fig:statisticalanalysis}
\end{figure*}

About 3950 $201\times201$ reconnection maps have been processed, which resulted in the detection of more than 254,000 individual spatiotemporal reconnection events. \textcolor{black}{For a driving period of $4$ days, this corresponds to about a reconnection event every second.} To investigate relative contributions of the events of different sizes, we constructed probability distribution histograms of $L$, $A$, $\tau$ and $V$ parameters (\autoref{fig:statisticalanalysis}). The distributions were computed using exponentially increasing bins to ensure uniform binning on the logarithmic scale. The event count in each bin was normalized to the bin width and the total number of events to obtain the normalized rate of occurrence. The power-law exponents of the probability distributions were estimated using a root-mean-square minimization of the power-law fit on the log-log scale.  \autoref{fig:statisticalanalysis} shows the resulting mean values and the standard errors of the exponents, with the dashed lines indicating the corresponding log-log slopes and the ranges of scales used for their calculation.

It can be seen that event linear sizes follow a very steep power law with slope $-3$ between the strand width of $1.5\times10^7\;\mathrm{cm}$ (i.e., a single reconnecting field line) and nearly $10^9\;\mathrm{cm}$, much larger than the diameter of a convective cell. The area of clusters of reconnecting strands follow a power law with slope $-2.8$ from about $10^5-10^{6.3}\;\mathrm{km^2}$. Typical events last for between $10^{2.5}-10^4\;\mathrm{s}$, and the lifetime distribution has a power law slope of $-2.6$. Finally, the spatiotemporal volume of events follows a power law slope of nearly $-2$. This analysis shows that there is no preferred spatial or temporal scale for nanoflares in the MHD simulation, but there is a clear collective behavior of the strands, with quite a steep fall off of event sizes. Evidently, the largest nanoflare clusters are bigger than the scale $a_0$ of the driver, and the longest lived clusters have a lifetime of the order of a rotation period of the driver, $\tau$. Furthermore, the majority of events have duration of about $200-300$ s, justifying our HD modeling assumption that events have a duration of $200$ s. The obtained scaling exponents are in agreement with, though slightly and uniformly steeper than, the scaling exponents describing spatiotemporal sizes of events found by using clusters of 2D features (current sheets, temperature changes, horizontal Poynting fluxes) as proxies for event sizes \citep{Knizhnik18a}. As demonstrated below, both exceedingly small and exceedingly large events are unlikely to produce the observed heating. Power laws are ubiquitious in the solar corona \citep[][and references therein]{Knizhnik18a}, and evidently the underlying reconnection also demonstrates frequent power laws.

\subsection{Hydrodynamics}
\subsubsection{Strands vs. Loops}\label{sec:strandVloop}

\begin{figure}
\includegraphics[width=\columnwidth]{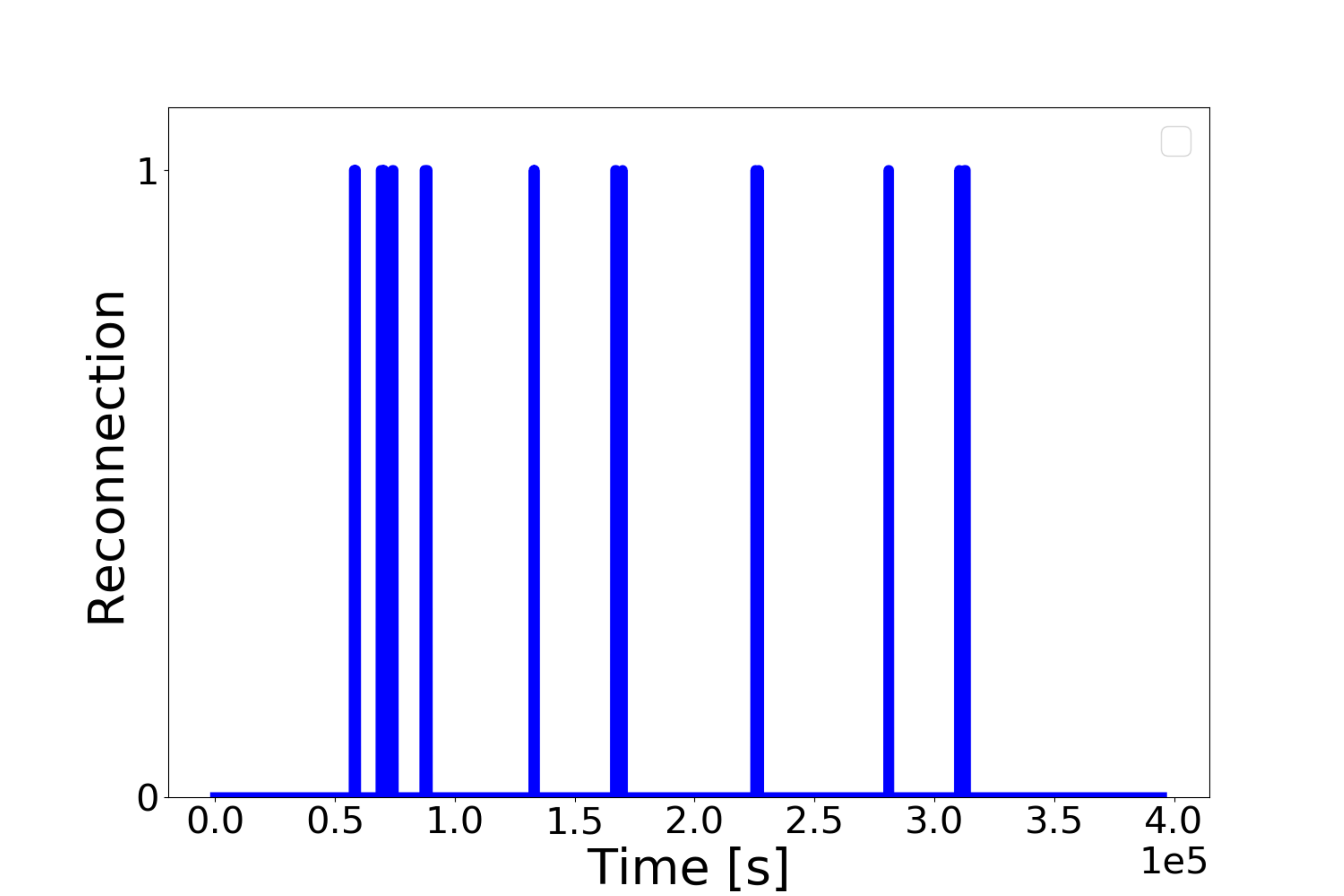}
\caption{The temporal behavior of a randomly chosen field line. $1$ represents a reconnection occurring, while $0$ represents ideal motion in response to photospheric driving.}
\label{fig:onesandzeroes}
\end{figure}

The time evolution of a given grid point corresponds to the temporal behavior of the field line traced from that seed point. If reconnection has occurred, it need not be considered the same field line, but it can nevertheless be associated with a given seed point. \textcolor{black}{In \autoref{fig:onesandzeroes}, we plot the time history of a field line (or, alternatively, seed point) chosen at random. The plot shows $1$ if reconnection has occurred, and $0$ if it has not (i.e., the field line moves only in response to photospheric driving). Since each reconnection can be associated with a heating event, or nanoflare, such temporal evolution of field lines in our simulation can be used to drive hydrodynamic models of a plasma's response to localized heating. In this example, the field line is considered as a `strand', and it is `heated' every time a reconnection event occurs. The field line's response to each heating event can be modeled using the EBTEL code, as described in \autoref{sec:ebtelsim}. Furthermore, we can combine multiple MHD strands to form a MHD `loop', by averaging the emission measures from multiple strands created this way. Since plasma is constrained to move along the field, in the absence of reconnection it cannot interact with other field lines, resulting in the possibility that a given cross section of a loop will have a range of temperatures \citep{Schmelz02,Schmelz14}. Therefore, in this study we follow the definition that ``a strand is an elementary flux tube where the physical properties of temperature and density are constant on the field lines'' \citet{Schmelz14}. To create a loop we calculate emission measure for each strand individually and then sum them as if all the strands were confined to a single observational pixel to obtain the $\dem$ of the entire loop.}

\subsubsection{EBTEL Simulations}\label{sec:ebtelsim}
\textcolor{black}{The primary inputs to EBTEL are the half-length of the loop and the time-dependent heating profile. We use a half-length of 10 Mm as the extent of the simulation box in the $x-$direction, $L_x=20 \;\mathrm{Mm}$, corresponds to the strand length (see \autoref{sec:initial_boundary_conditions}). Additionally, we use a flux-limiting coefficient of $1/6$ to compensate for artificially increased conductive cooling at times when the density cannot support classical Spitzer-H\"{a}rm thermal conduction \citep{Spitzer53}. For additional details, see \citet[Section 2.1]{Barnes16}. To calculate the initial temperature and density, we impose a constant background heating rate of $q_{\text{bg}}=1.96\times10^{-7}$ erg cm$^{-3}$ s$^{-1}$ and solve the EBTEL equations in hydrostatic equilibrium such that the initial temperature and density are approximately $0.115$ MK and $5.46\times10^6$ cm$^{-3}$, respectively. This background heating is sustained throughout the entire simulation to ensure that the strand does not cool to negative temperatures.}\par 
\textcolor{black}{We simulate the thermodynamic evolution of each traced strand for the entire ARMS simulation period, approximately 4.5 days. To derive the nanoflare heating profile from the results of the MHD simulation, we assume that each identified discrete reconnection event corresponds to a single nanoflare heating event. Each nanoflare has a symmetric triangular heating profile of total duration 200 s and a peak heating rate of $q$ such that at the onset of the event, the heating rate ramps up linearly from $q_{\text bg}$ to $q$ over a period of 100 s and then ramps linearly back down to $q_{\text bg}$ over the last 100 s. A fixed event duration of 200 s allows the energy release to be relatively impulsive compared to the cooling time scale of the loop (on the order of $10^3$ s). Furthermore, we choose to deposit all of the energy in the electrons. Though we could have chosen to deposit some portion of energy in the ions as well, the partition between the two species is not likely to impact observable signatures of interest as the two fluids have already mostly equilibrated by the time they reach the radiative cooling phase of their evolution \citep{Barnes16,Barnes16b}.}
\par 
\textcolor{black}{Although the MHD model conserves energy, obtaining the energy released by each event is complex. Several studies have obtained energy release distributions by integrating $\eta\textbf{J}^2$ over the current sheets in the system \citep{Kanella17,Kanella18,Kanella19}, but it has been argued that energy release occurs via viscous, rather than Ohmic, dissipation \citep{Knizhnik19}.\footnote{\textcolor{black}{In our model, the viscosity is numerical, rather than explicit, and as a result is highly spatially- and temporally-varying. Thus, quantifying the viscous heating is extremely challenging.}} \textcolor{black}{In any case, evaluating the Ohmic heating may overestimate the energy of each event, which likely converts only a fraction of the available magnetic energy into heating. In order to avoid such uncertainties, as a first step,} therefore, we assume that each event has an energy that depends on the time since the previous event \citep{Cargill14} such that the peak heating rate for event $i$ on a given strand is,
\begin{equation}\label{eq:c14heating}
q_i = q_{\text{max}}\frac{t_{\text{wait},i}}{t_{\text{wait},\text{max}}}
\end{equation}
where $t_{\text{wait},i}$ is the time interval preceding event $i$, $t_{\text{wait},\text{max}}$ is the maximum time interval for a given strand, and $q_{\text{max}}$ is chosen such that the emission measure distribution (see \autoref{sec:em_diagnostic}) peaks at approximately 4 MK, consistent with observations of active region cores \citep[e.g.,][]{Warren12}.}

\subsubsection{EBTEL Results: A Single Strand}\label{sec:singlestrand}

\begin{figure*}
\includegraphics[width=0.5\linewidth]{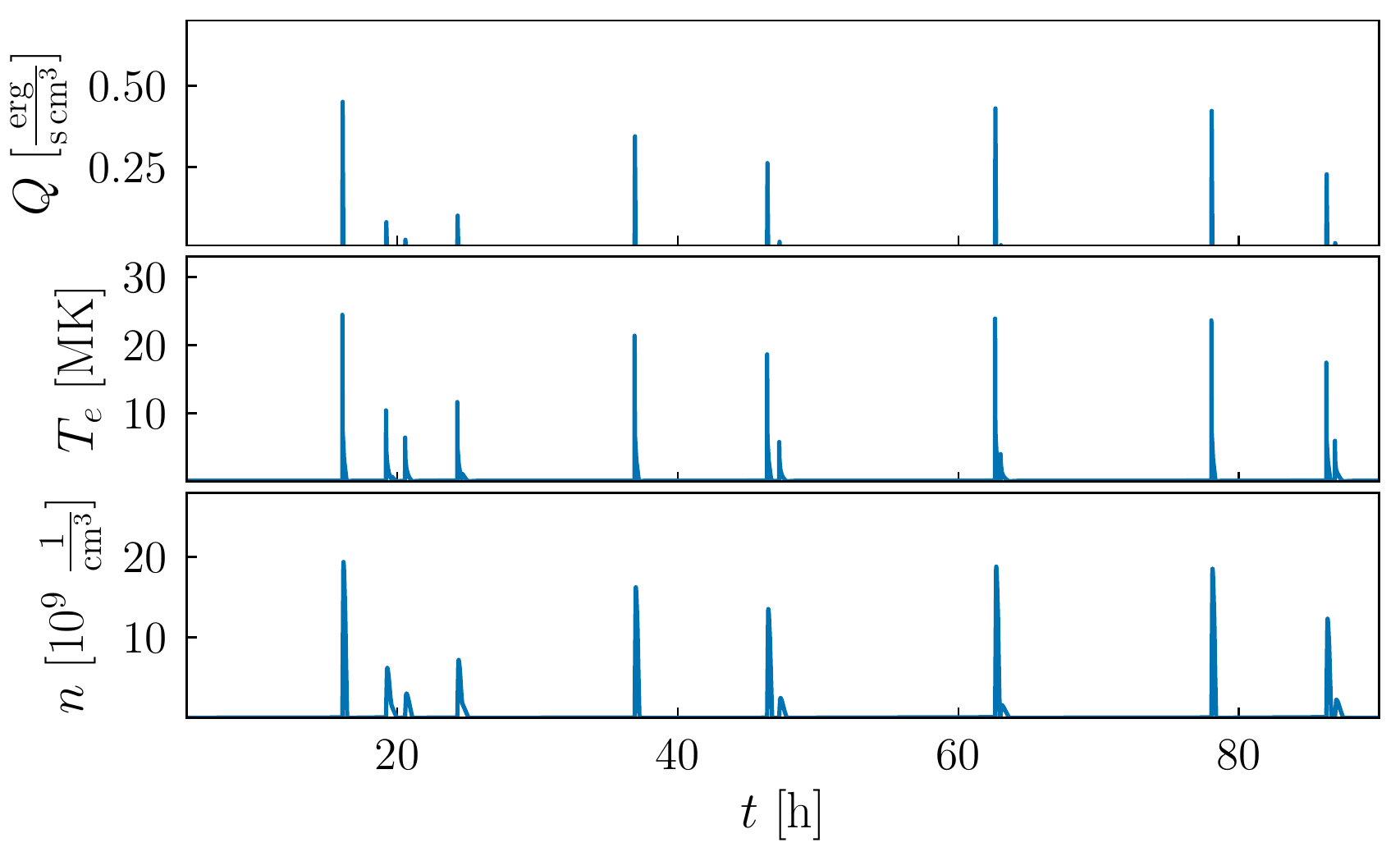}
\includegraphics[width=0.5\linewidth]{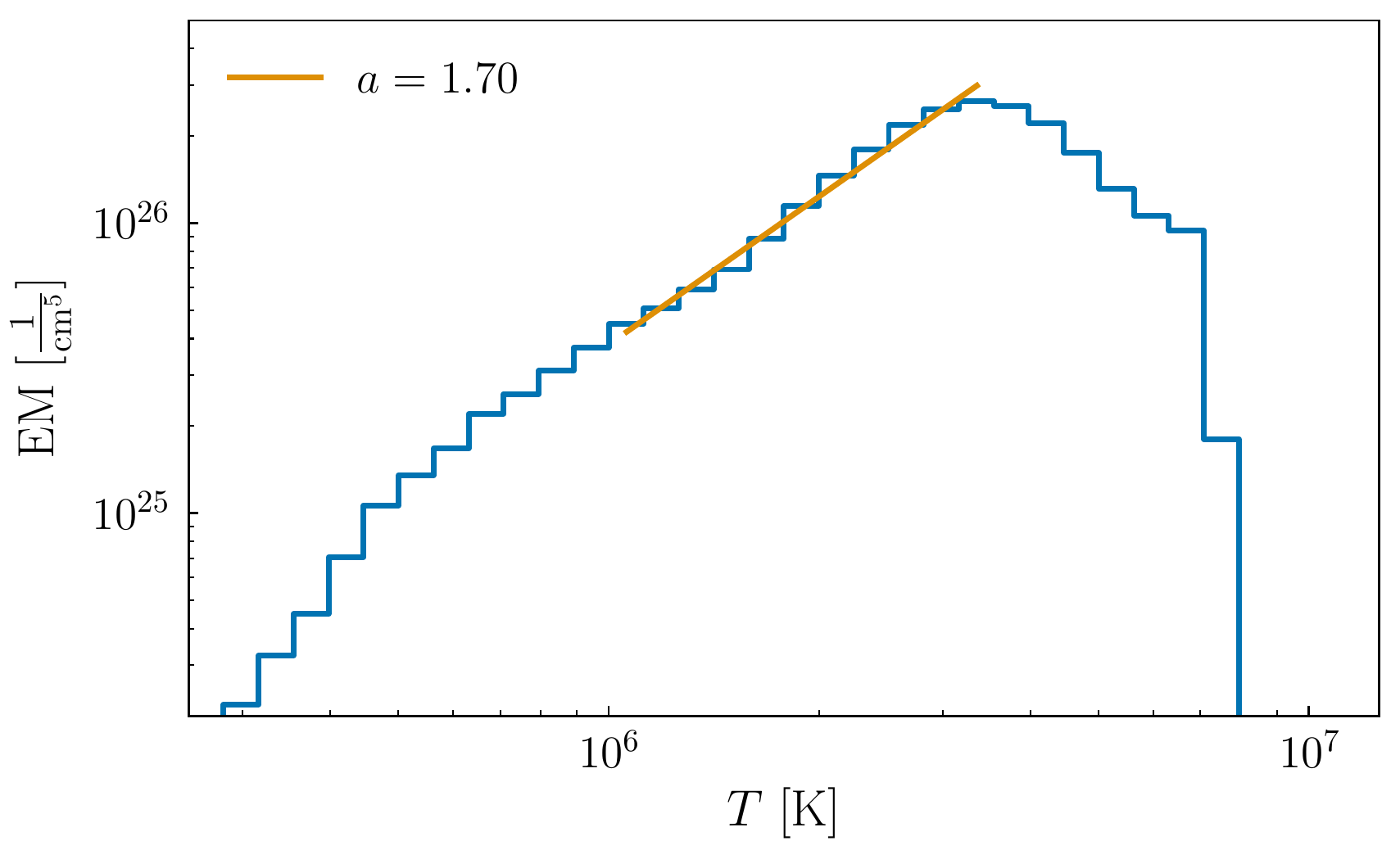}
\caption{\textbf{Left:} Heating rate (top), electron temperature (middle), and density (bottom) as a function of time for the period $5<t<90$ h as computed by EBTEL for the same strand shown in \autoref{fig:onesandzeroes}. \textbf{Right:} Resulting $\dem$ distribution time-averaged over the whole simulation period. The overlaid orange line shows the power-law fit between $10^6$ K and $T_M$, the peak of the distribution, and the resulting slope, $a$, from the fit is given in the legend.}
\label{fig:onestrand}
\end{figure*}

For each of the $201^2$ traced strands in the red box in \autoref{fig:InstRx}, we ran a two-fluid EBTEL simulation for the entire ARMS simulation period to compute the time-dependent temperature and density evolution in response to the reconnection-derived heating profiles. \autoref{fig:onestrand} shows the hydrodynamic evolution (left) and  time-averaged $\dem$ distribution (right) for \textcolor{black}{the strand shown in \autoref{fig:onesandzeroes}}. As can be seen from the temperature (middle left) and density (bottom left) profiles, the cooling and draining of the strand occurs on a much shorter timescale than the period between consecutive heating events (top left) in nearly all cases. The resulting $\dem$ (right) has a shallow low-temperature slope (denoted by the orange line), consistent with low-frequency nanoflare heating \citep[e.g.][]{Cargill14}. 

\begin{figure*}
\includegraphics[width=0.5\linewidth]{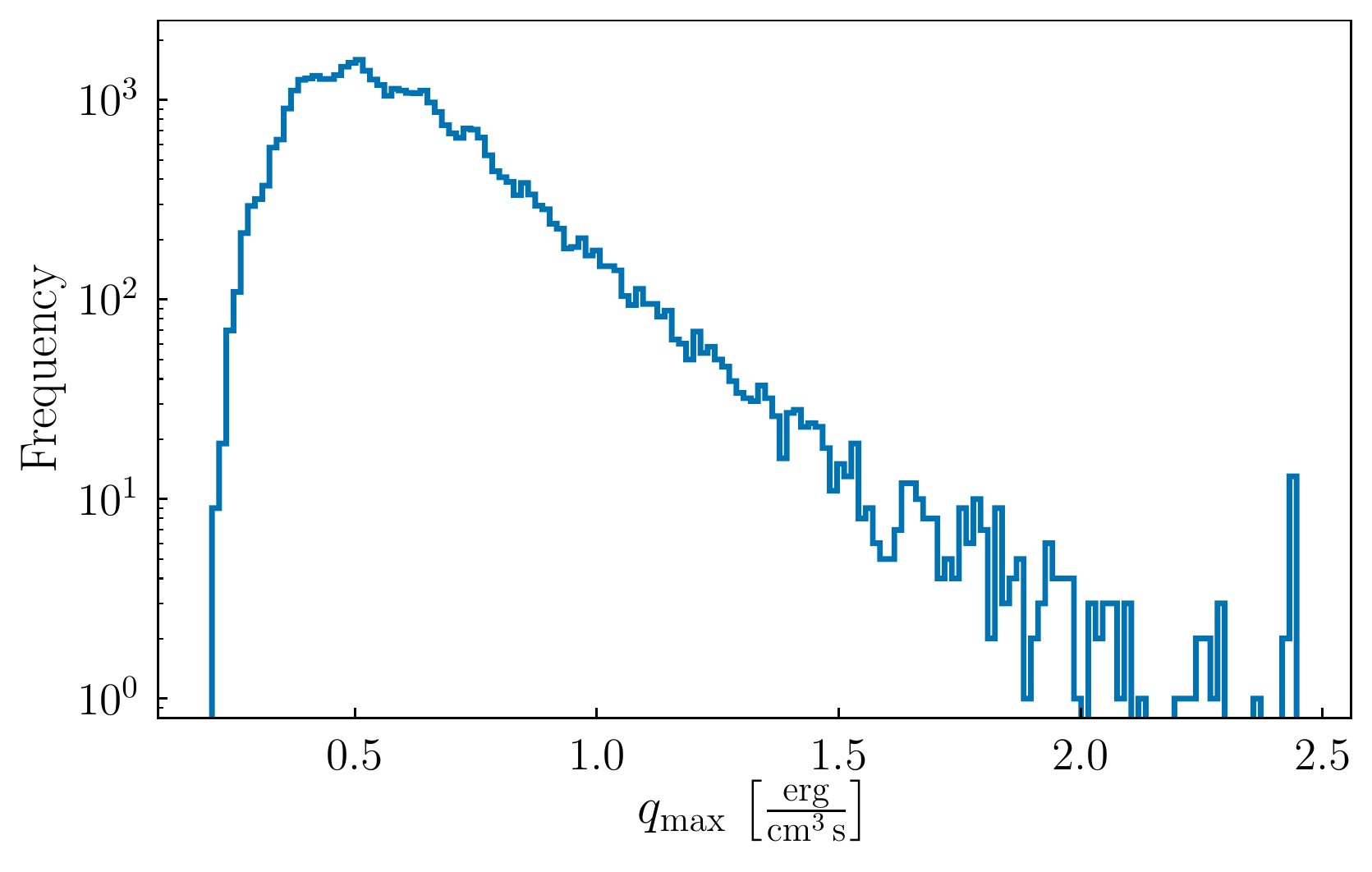}
\includegraphics[width=0.5\linewidth]{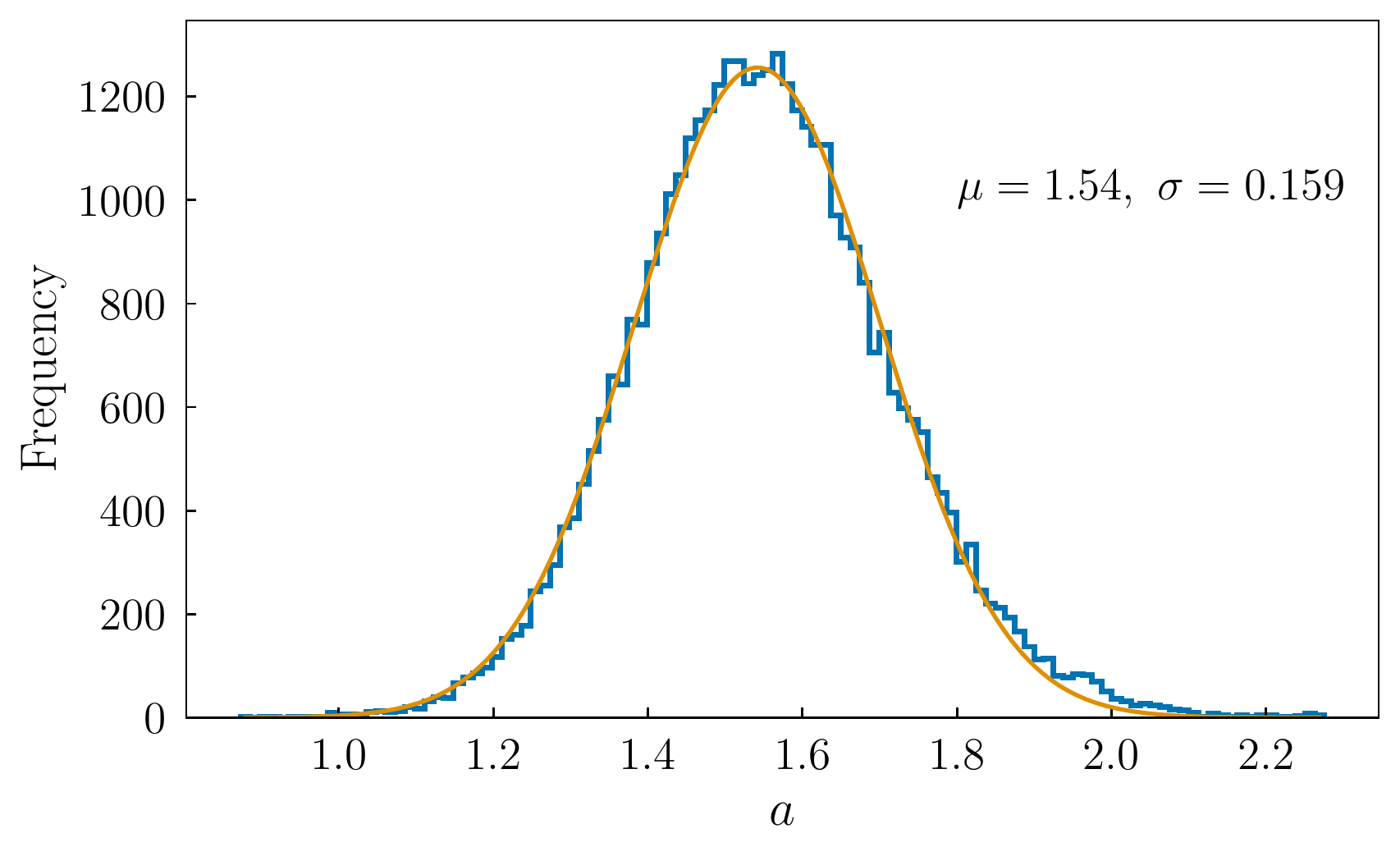}
\caption{\textbf{Left:} Distribution of maximum heating rates, $q_{\text max}$, for each traced strand. $q_{\text max}$ is the heating rate corresponding to the event with the longest delay time preceding it and its value is chosen such that the resulting $\dem$ distribution has $T_M\approx4$ MK. The heating rates of all the other events for a given strand scale with $q_{\text max}$ according to \autoref{eq:c14heating}. \textbf{Right:} The blue histogram denotes the distribution of slopes from the $\dem$ distribution from each of the $201^2$ strands. The orange curve shows the Gaussian fit to the distribution and the resulting mean, $\mu$, and standard deviation, $\sigma$, are indicated in the plot. The uncertainties in both the mean and standard deviation as derived from the Gaussian fit are of order $10^{-3}$.}
\label{fig:allstrandsindividual_max_heating_dist}
\end{figure*}

The left panel of \autoref{fig:allstrandsindividual_max_heating_dist} shows the distribution of maximum heating rates, $q_{\text max}$, for all $201^2$ strands. $q_{\text max}$ corresponds to the event on that strand with longest delay time and the heating rate of the remaining events are scaled according to \autoref{eq:c14heating}. The peak heating rate is of the order of $0.5\;\mathrm{erg\;cm^{-3}\;s^{-1}}$ for most strands. Additionally, the right panel of \autoref{fig:allstrandsindividual_max_heating_dist} shows the distribution of $\dem$ slopes, $a$, as calculated from the $\dem$ distributions for all $201^2$ strands. The slope distribution is approximately Gaussian, with a mean and standard deviation of $1.54$ and $0.16$, respectively\textcolor{black}{. This distribution of model slopes, produced by low frequency nanoflares resulting from the driving pattern in the MHD simulation, cannot account for the full range of observed emission measure slopes, $1.7\lesssim a\lesssim5$ \citep{Tripathi11,Warren11,Warren12,Winebarger11}.}


\subsubsection{EBTEL Results: Clusters of Strands}
\label{sec:ebtelcombined}

\begin{figure*}
\includegraphics[width=0.5\linewidth]{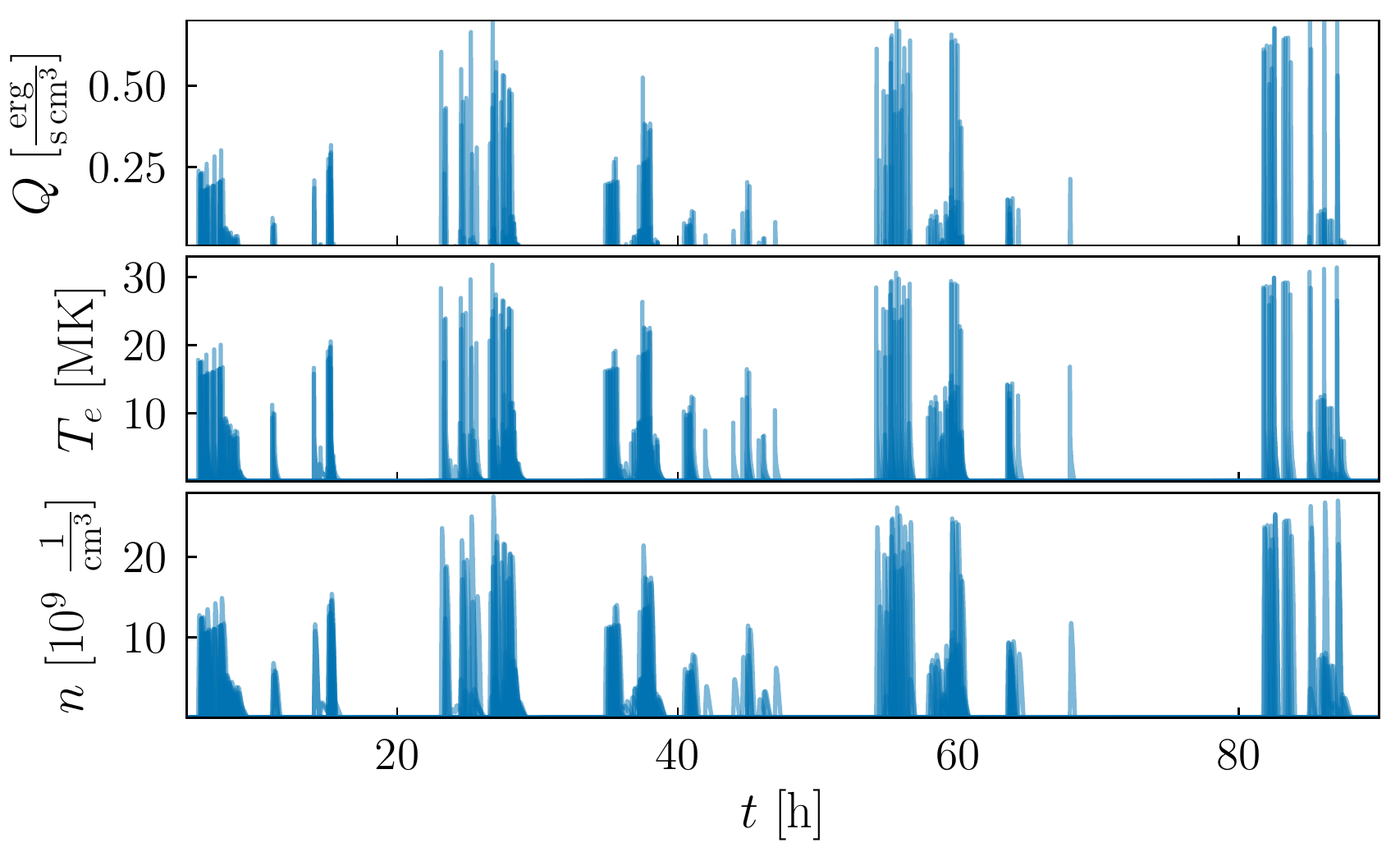}
\includegraphics[width=0.5\linewidth]{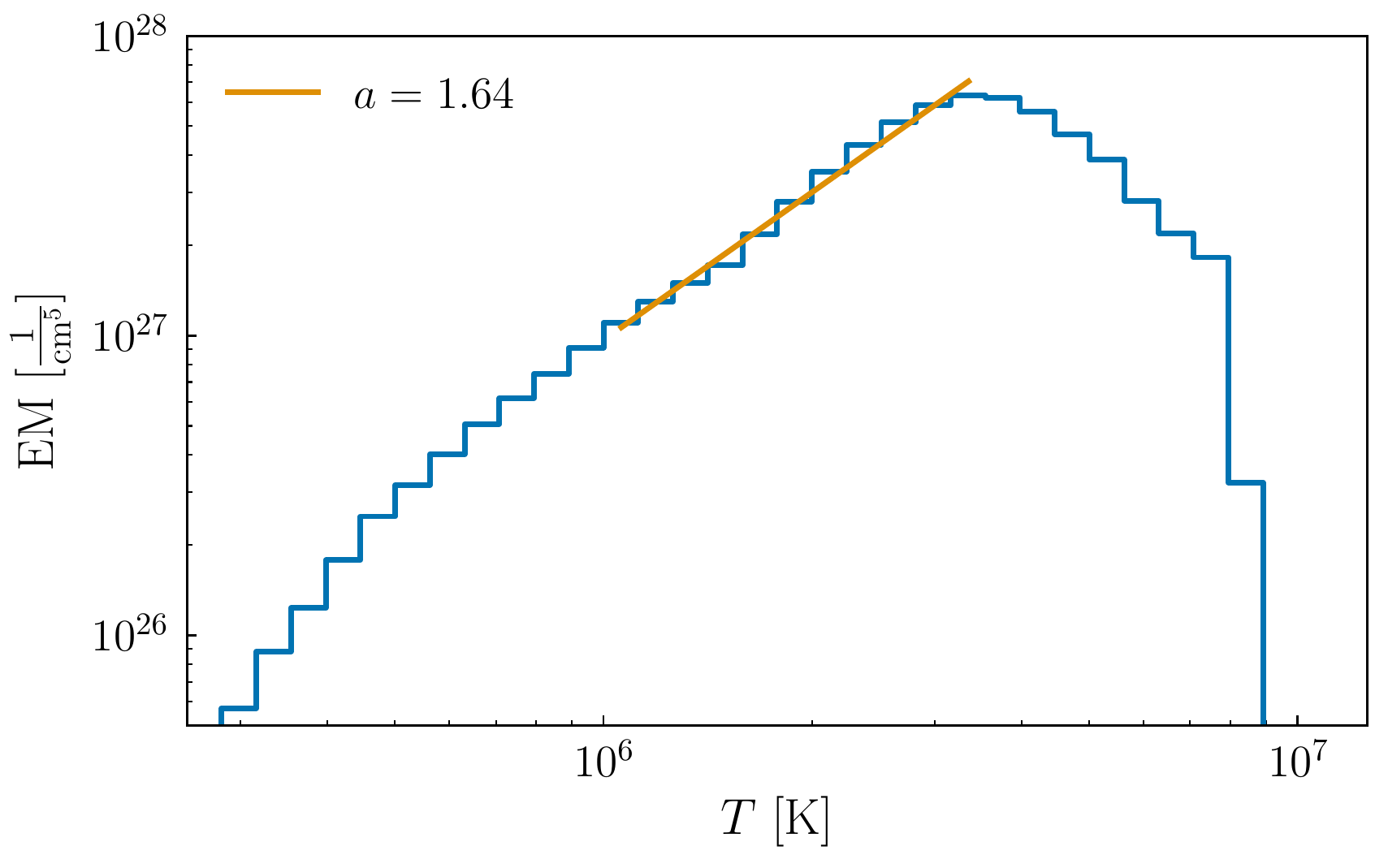}
\caption{\textcolor{black}{\textbf{Left:} Same as \autoref{fig:onestrand}, but now the heating, temperature, and density profiles for all 25 strands from the subgrid are overlaid on a single plot. \textbf{Right:} Summation of all 25 $\dem$ distributions from the strands in the subgrid. The summed $\dem$ distribution is then time-averaged over the simulation period.}}
\label{fig:subset}
\end{figure*}

We next consider a $5\times5$ randomly chosen subgrid inside the red box in \autoref{fig:InstRx}, \textcolor{black}{such that each of the 25 strands is an unresolved elemental flux tube comprising a single loop}. Such a loop would have a width of about $750\;\mathrm{km}$, approximately the size of loops as measured by the \emph{Hi-C} rocket \citep{Brooks13,Aschwanden17}. The left panel of \autoref{fig:subset} shows the time-dependent heating rate, temperature, and density for all 25 strands. \textcolor{black}{Summing over all 25 unresolved elementary strands, the loop is heated by a total of 1415 events over the whole simulation period. Note that the events tend to cluster in groups such that they resemble nanoflare ``storms,'' sequences of nanoflares that repeat in quick succession, but on different strands such that each strand is heated approximately once during each storm \citep{Klimchuk15}.}

The right panel of \autoref{fig:subset} shows the resulting $\dem$ distribution, \textcolor{black}{computed by summing the $\dem$ distributions for all $25$ strands in the subgrid and then time-averaging over the simulation period}. The resulting slope for the loop composed of 25 strands is \textcolor{black}{quite similar to the slopes of the single strand case, falling less than 1 standard deviation above the mean for the single strand slope distribution shown in the right panel of \autoref{fig:allstrandsindividual_max_heating_dist}.}

\begin{figure*}
\includegraphics[width=0.5\linewidth]{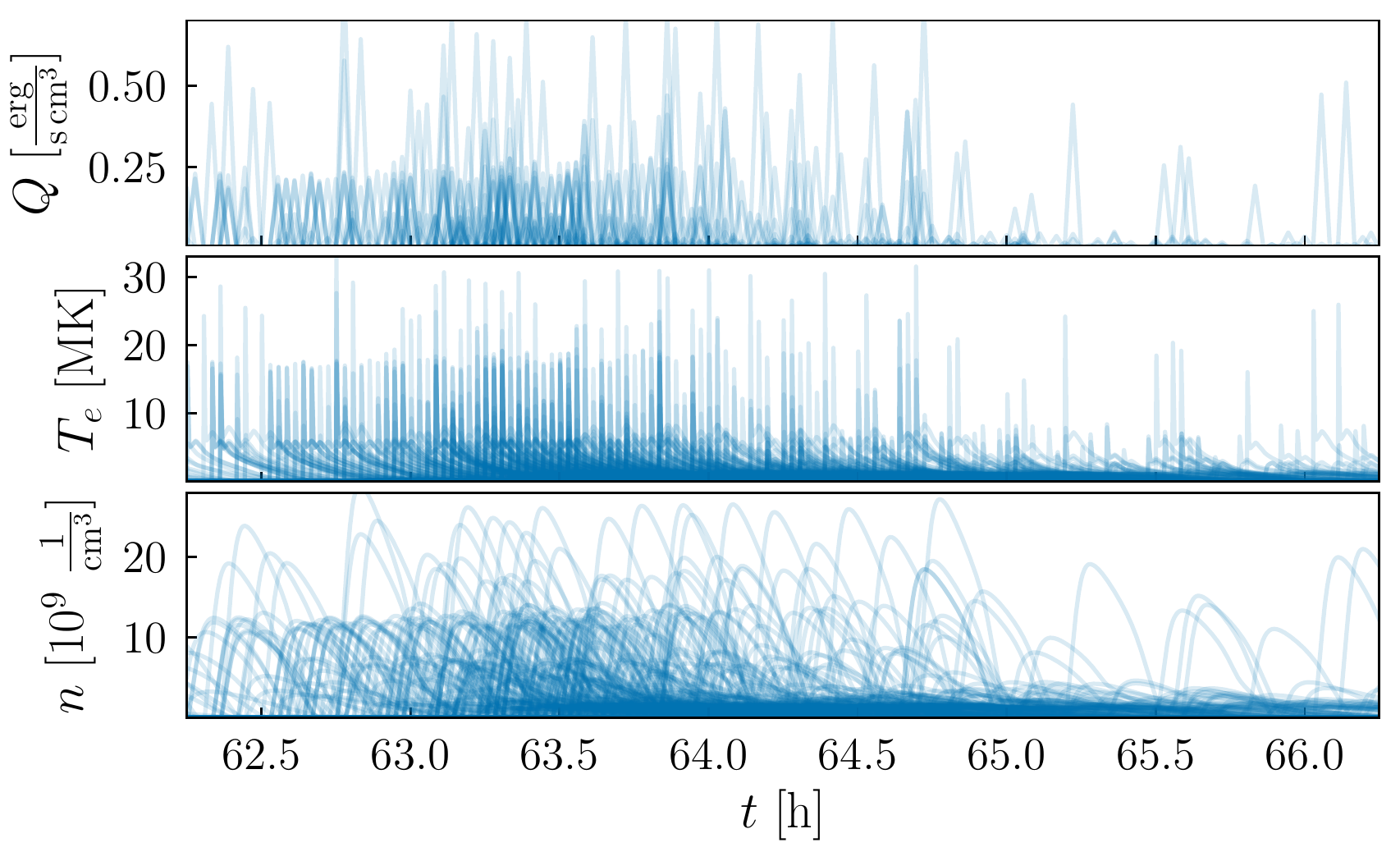}
\includegraphics[width=0.5\linewidth]{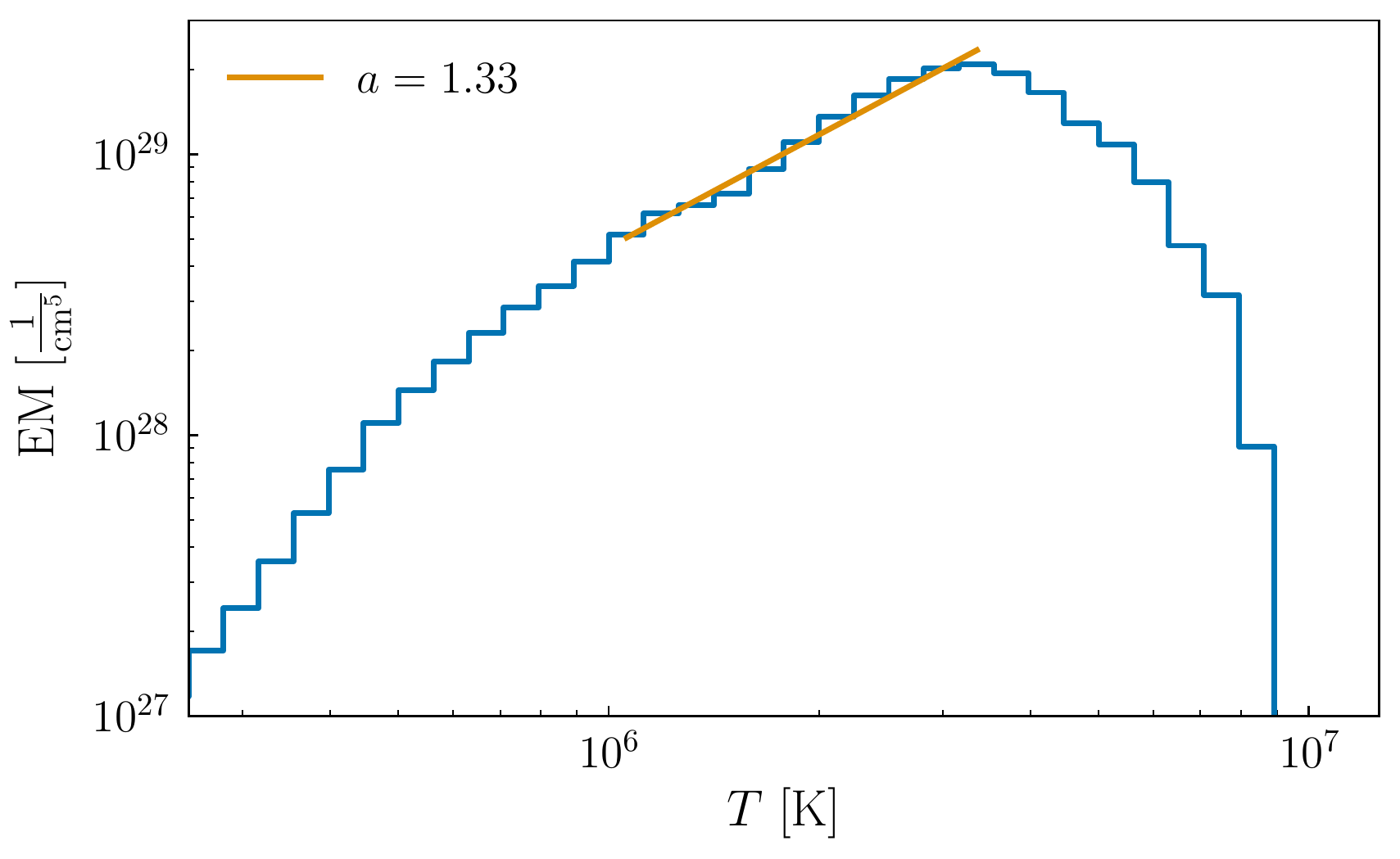}
\caption{Same as \autoref{fig:subset}, but in the case of a loop comprised of the $353$ strands identified by the clustering algorithm. In this case, the $\dem$ distribution is time-averaged only over that time interval covered by the identified cluster.}
\label{fig:clusterstrands}
\end{figure*}

We apply the clustering algorithm described in \autoref{sec:collective} to identify groupings of strands in space and time. 
For one such identified cluster composed of $353$ strands, we apply the same procedure as above, simulating the evolution of a loop \textcolor{black}{comprised of all strands in the cluster}.
In this experiment, the `loop' is not simply a square subgrid of adjacent seed points, but is a collection of contiguous field lines behaving collectively. The resulting hydrodynamic evolution and \textcolor{black}{time-averaged} $\dem$ distribution are shown in \autoref{fig:clusterstrands}.
We simulate the evolution of the loop over only that time interval during which the strands are behaving collectively \textcolor{black}{as identified by the clustering algorithm.
Here, too, the slope of the total time-averaged $\dem$ is relatively shallow, showing a broad distribution of temperatures.
The slope falls slightly more than 1 standard deviation below the mean of the single strand slope distribution in \autoref{fig:allstrandsindividual_max_heating_dist}, consistent with very low-frequency nanoflare heating and not dissimilar from the single strand cases.}

\subsubsection{EBTEL Results: All Strands}

\begin{figure*}
\includegraphics[width=0.5\linewidth]{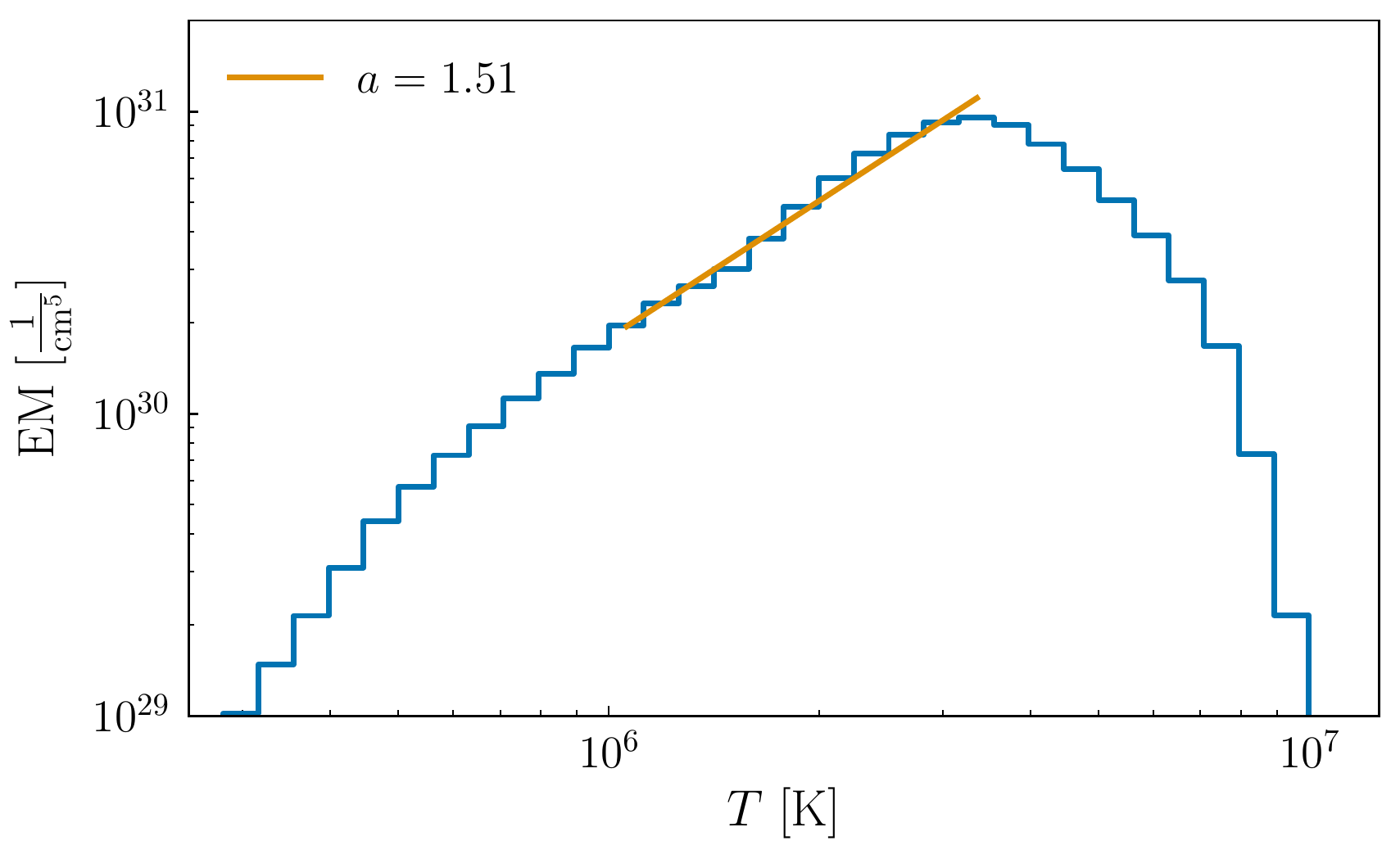}
\includegraphics[width=0.5\linewidth]{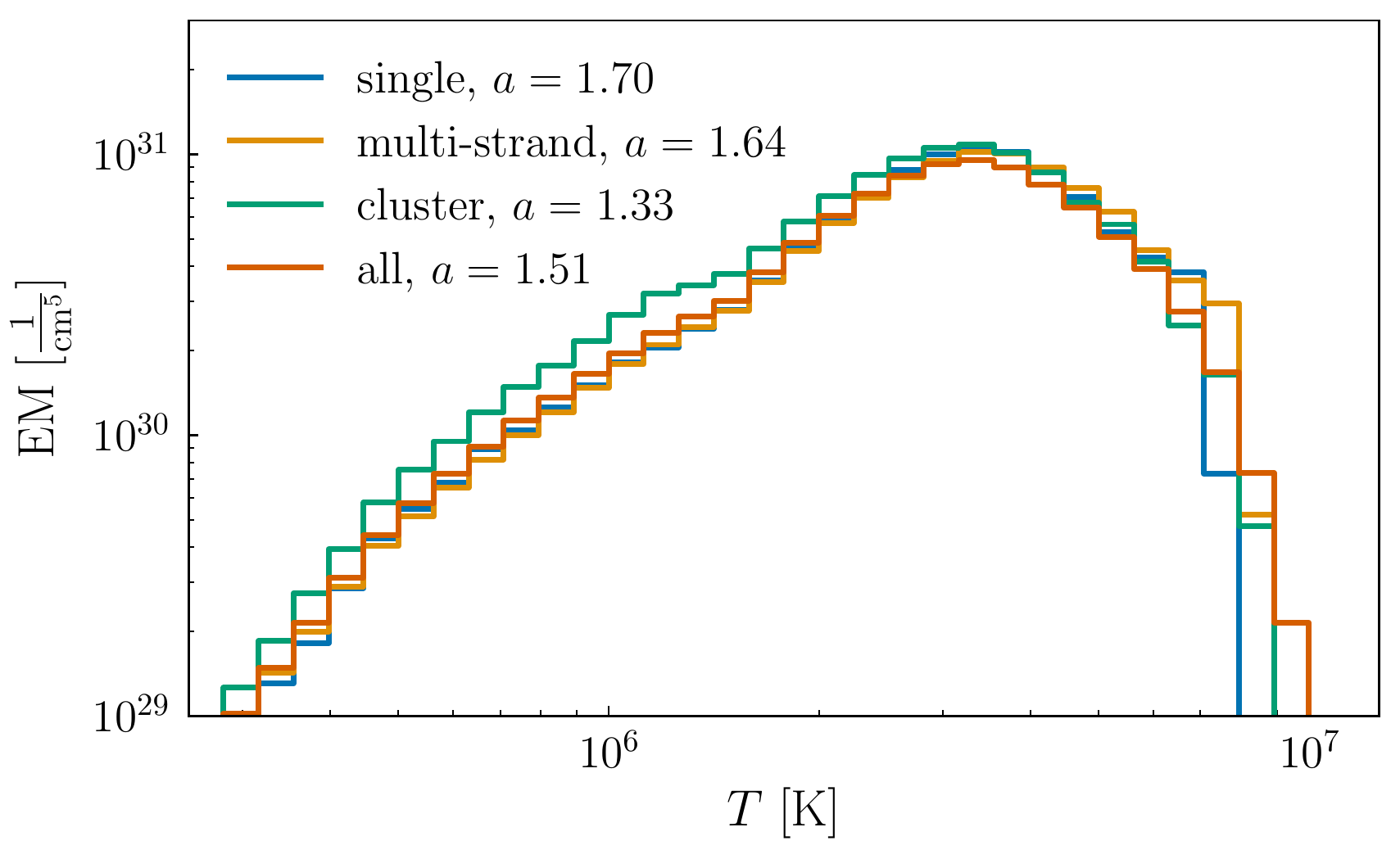}
\caption{\textbf{Left:} Same as the right panels of \autoref{fig:subset} and \autoref{fig:clusterstrands}, but for the case of a loop comprised of all $201^2$ strands in the red box in \autoref{fig:InstRx}. \textbf{Right:} Comparison between the $\dem$ distributions for each case. The single, subgrid, and cluster distributions have been scaled up such that the emission at the peak of the distribution in each case is equal to the all-strands case. The slope for each case is denoted in the legend.}
\label{fig:allstrands}
\end{figure*}

Finally, we consider all $201^2$ strands inside the red box in \autoref{fig:InstRx} as comprising a single `loop' of width $3\times10^9\;\mathrm{cm}$. \textcolor{black}{We compute one $\dem$ by binning the temperatures, weighted by the density squared, of all $201^2$ loops. The resulting time-averaged $\dem$, is shown in the left panel of \autoref{fig:allstrands}. As with the previous cases, the measured cool $\dem$ slope, denoted in the legend of the left panel of \autoref{fig:allstrands}, is not significantly different from the single strand case. In particular, the slope for the loop comprised of all $201^2$ strands is less than one standard deviation below the mean of the distribution of single strand slopes (see right panel of \autoref{fig:allstrandsindividual_max_heating_dist}).}

\textcolor{black}{The right panel of \autoref{fig:allstrands} shows a comparison between the $\dem$ distributions for the single-strand, subgrid, cluster, all-strands cases. The first three cases have been scaled such that they are approximately equal to the all strands case at the peak of the distribution. The slopes for each case are denoted in the legend. While the magnitudes of each distribution vary greatly due to the number of strands, and thus the amount of emitting material, included in the loop, there is little variation in the shape of the $\dem$ distributions. This is partially captured by the slope parameter.}

\section{Conclusions and Implications}
\label{sec:implications}

In this paper, we presented \textcolor{black}{results from HD models} driven by heating profiles derived from an MHD simulation. To our knowledge, this is the first model to feed MHD results directly into HD simulations. We simulated a \citet{Parker72} corona driven by complex photospheric motions that twist and braid the magnetic field to follow the footpoint motion of individual magnetic field lines and identify reconnection frequencies. We then used the behavior of both individual and groups of field lines to obtain $\dem$ \textcolor{black}{distributions} from \textcolor{black}{our} HD modeling \textcolor{black}{results}. Our main results can be summarized as follows: 
\begin{enumerate}
    \item The distribution of nanoflare waiting times follows a power law distribution with a slope of about $-1.3$, slightly steeper than the more simplified model of \citet{Knizhnik20}, but far shallower than previously assumed \citep{Cargill14,Bradshaw16,Barnes16b,Barnes19}.
    \item Reconnection on a single strand of width $150\;\mathrm{km}$ \textcolor{black}{produces a narrowly peaked distribution of $\dem$ slopes that is well below the range of slopes derived from active region core observations. Thus, using the $\dem$ slope as a proxy for the nanoflare heating frequency, the MHD driver considered here cannot fully account for the range of heating frequencies observed in active region cores.}
    \item \textcolor{black}{Combining multiple strands into a loop and computing the \textcolor{black}{total time-averaged} $\dem$ does not significantly change the resulting $\dem$ distribution, with the slopes of nearly all other cases falling within 1 standard deviation of the mean of the single strand slope distribution. This indicates that the photospheric driving of the field is approximately uniform throughout the region \textcolor{black}{and that combining strands that are heated out of phase does not significantly alter our conclusions about the underlying heating frequency based on the shape of the $\dem$ distribution.}}
    \item Field lines in the MHD model behave collectively, and reconnect in clusters. The width, area, and lifetimes of these clusters follow power laws, each with slopes of between $-3.0$ and $-2.5$. The obtained statistics suggest an effective interaction between individual nanoflares across a wide range of spatial scales, up to the system-size scale imposed by the diameter of the simulated loop system.
    \item While there is no preferred spatial scale for these clusters, the largest clusters are bigger than the length scale of the driver.
    \item While there is no preferred temporal scale for these clusters, the longest lived clusters have a lifetime of the order of a rotation period of the driver.
\end{enumerate}
\par 
\textcolor{black}{The collective behavior seen in our model and displayed in \autoref{fig:statisticalanalysis} suggests that loops are the result of a random clumping of energization that lights up a group of neighboring field lines at once. This could also correspond to the diffuse emission between loops, which can also be bright and behave similarly to the loops themselves \citep{DelZanna03}.} \par 
\textcolor{black}{The results shown in  \autoref{fig:onestrand}-\autoref{fig:allstrands} demonstrate that the $\dem$ slopes derived from our model are more or less independent of the number of strands that comprise a loop, because the dynamics of each individual strand are quite similar. No matter how many strands we include in a loop, our measured slopes cannot account for the range of observed $\dem$ slopes, typically fall between $2$ and $5$, \citep[e.g.,][]{Tripathi11,Warren11,Winebarger11,Schmelz12,Warren12,DelZanna15}. Although our measured slopes are not consistent with observations, it should be noted that error bars on measured slopes are not insignificant. The chosen temperature bounds of the fit \citep{Barnes16b}, the inversion method used to calculate the $\dem$ \citep{Warren17}, or uncertainties in the atomic data \citep{Guennou13} can all play a role in the nominal value and error bars of the slope. }
\par 
\textcolor{black}{The choice of loop half length is an important parameter in our analysis. Our choice of loop length, $L_x=20$ Mm, corresponds to short, core active region loops \citep{Reale10}. This has critical consequences for the response of the plasma, since the cooling time of a loop depends on its length, (\citet{Cargill94} and \citet{Cargill14} find that the cooling time is related to loop length as $L^{5/6}$) i.e. longer loops cool slower than shorter loops, and the nanoflare frequencies such as those measured here could result in high frequency heating on longer loops - for the same reconnection and heating rates - due to their corresponding cooling times. It will therefore be important to understand the response of the plasma to magnetohydrodynamic heating profiles for a range of loop lengths}
\par 
\textcolor{black}{While we find that the observables derived from our MHD-driven HD models are not consistent with the full range of active region core observations,} our work combines two previously disconnected types of analyses: realistic treatments of magnetic field behavior and realistic treatments of plasma response to heating. However, further work is necessary to ascertain whether the peak heating rates used in our model are well-representative of those in simulations. One challenge of our MHD model is that it is difficult to obtain heating rates directly from the individual MHD reconnection events. The reasons for this were described in more detail in \citet{Knizhnik20}, but addressing this question is of vital importance for fully combining the MHD and HD models. This work was able to reduce the number of \emph{ad hoc} inputs to HD models from two (nanoflare time and energy distributions) down to one (just the energy distribution). Obtaining physically motivated MHD values for the energy distribution to feed into the HD models will be the subject of future work.
\par
\textcolor{black}{One possible drawback to our model is that the feedback from the plasma density and temperature changes on the reconnection physics is neglected. Changes in the plasma parameters will affect both the local Alfv\'en speed (and thus the reconnection rate), and local plasma $\beta$. These changes could have important consequences for the frequency of subsequent reconnection events. Further investigation is required to address this issue.}
\par 
We conclude by addressing the question of the numerical resolution's effects on our results. In our simulation, our finite numerical dissipation allows reconnection to occur at electric current sheets associated with discontinuities in the direction of the magnetic field. As a result, the scales of magnetic reconnection are set by the grid resolution and it is legitimate to ask whether increasing the resolution of our simulation would fundamentally alter our results. \citet{Knizhnik20} performed their analysis with both explicit and numerical resistivity, the former removing the dependence on resolution by allowing reconnection to occur via a diffusive term in the MHD momentum equation. They obtained results that were qualitatively and quantitatively similar in the two cases. \citet{Knizhnik19} performed an analysis of heating resulting from photospheric driving very similar to that presented here at multiple different resolutions. They found that the heating rate increased with the logarithm of the Lundquist number, (i.e., the inverse of the effective resistivity) so that extrapolating to coronal resistivities would increase the simulated heating rate by only about a factor of $10$. As a result, we can conclude that our chosen grid resolution is not significantly affecting our results. \par

\acknowledgments
K.J.K.\ and V.M.U.\ were supported by NASA GSFC's Internal Scientist Funding Model (competitive work package on “Understanding Coronal Heating and the Solar Spectral Irradiance"). They would like to thank Jim Klimchuk, the lead of the work package team, for fruitful discussions. K.J.K.\ was also supported by the Jerome and Isabella Karle Distinguished Scholar Fellowship and the Office of Naval Research. W.T.B.\ was supported for this work by NASA's \textit{Hinode} program through the National Research Council.  J.W.R. was supported by NASA's \textit{Hinode} program. The authors are grateful to the anonymous referee for important feedback that greatly improved the quality and scope of the work.
\par 

\software{
Astropy \citep{Astropy18},
Dask\citep{Dask15},
IPython \citep{IPython07},
Matplotlib \citep{Matplotlib07},
Numpy \citep{Numpy06},
Scipy \citep{Scipy20},
seaborn \citep{Seaborn18},
Zarr\citep{Zarr20}
}
\newpage 
\bibliographystyle{aasjournal}
\bibliography{bibliography}

\end{document}